\documentclass[12pt,preprint]{aastex} 

\newcommand{\degree}{^{\circ}}

\newcommand{\solarmass}{M_{\odot}}

\def\3EG{{3EG J1746-2851}}
\def\p0{{$\pi^0$}}
\def\1018{{$10^{18}$}}
\def\spin{{spectral index }}
\def\norm{{normalization }}

\newcommand{\be}{\begin{equation}}
\newcommand{\ee}{\end{equation}}
\newcommand{\bea}{\begin{eqnarray}}
\newcommand{\eea}{\end{eqnarray}}
\newcommand{\bmu}{\begin{multline}}
\newcommand{\emu}{\end{multline}}

\def\msun{{\,M_\odot}}

\def\simlt{\lower.5ex\hbox{$\; \buildrel < \over \sim \;$}}
\def\simgt{\lower.5ex\hbox{$\; \buildrel > \over \sim \;$}}

\def\gcm3{{\rm\,g\,cm^{-3}}}
\def\ncm3{{\rm\,cm^{-3}}}

\def\>{$>$}
\def\<{$<$}

\lefthead{Crocker et~al.}                      
\righthead{Sgr B Cosmic Rays}

\received{}
\begin{document}


\title{The Cosmic Ray Distribution in Sagittarius B}

\renewcommand{\thefootnote}{\fnsymbol{footnote}}

\author{Roland M. Crocker$^{1}$, David Jones$^{1,2}$, Raymond J. Protheroe$^{1}$, 
J{\" u}rgen Ott$^{2,6}$\footnote{Currently: Jansky Fellow, National Radio Astronomy Observatory
520 Edgemont Road,
Charlottesville, VA 22903, email: jott@nrao.edu}, Ron Ekers$^{2,7}$,
Fulvio Melia$^{3,8}$, Todor Stanev$^{4}$, and Anne Green$^{5}$}
\affil{$^1$ School of Physics and Chemistry\\
University of Adelaide\\
5005, Australia\\
roland.crocker@adelaide.edu.au\\
djones,rprother@physics.adelaide.edu.au}

\affil{$^2$ Australia Telescope National Facility\\
Marsfield,\\
2122, Australia\\
Juergen.Ott,Ron.Ekers@csiro.au}

\affil{$^{3}$Physics Department and Steward Observatory, 
\\The University of Arizona, Tucson, AZ 85721\\
melia@physics.arizona.edu}

\affil{$^{4}$Bartol Research Institute, \\ 
The University of Delaware, Newark, DE 19716-2593\\
 stanev@bartol.udel.edu}

\affil{$^{5}$School of Physics, \\
University of Sydney, \\
2006, Australia \\
agreen@physics.usyd.edu.au}

\affil{$^{6}$Bolton Fellow}

\affil{$^{7}$Federation Fellow}

\affil{$^{8}$Sir Thomas Lyle Fellow and Miegunyah Fellow}

\date{\today}

\newpage

\begin{abstract}

The HESS instrument has observed a 
diffuse 
flux of $\sim$ TeV $\gamma$-rays 
from a large solid angle  
around the Galactic center (GC).
This emission is 
correlated with the distribution of gas in the region suggesting that
the  $\gamma$-rays  originate in 
collisions between
 cosmic ray hadrons (CRHs)  
and ambient matter.
Of particular interest, HESS has detected
$\gamma$-rays from
the Sagittarius (Sgr) B Molecular Cloud Complex. 
Prompted by the suggestion of a hadronic origin for the gamma rays, 
we have  examined archival 330 and 74 MHz  Very Large Array radio
data and 843 MHz 
Sydney University Molonglo Sky Survey data
covering Sgr B,
looking for synchrotron
emission 
from secondary electrons and positrons 
(expected to be created
in the same interactions 
that supply the observed
gamma rays). 
Intriguingly, 
we have uncovered non-thermal emission, but at a level exceeding expectation.
Adding to the overall picture,
recent observations by the Atacama Pathfinder Experiment 
telescope
show that the cosmic ray ionization rate 
is ten times greater in the Sgr B2 region of Sgr B  than the local value.
Lastly, Sgr B2 is also a very bright X-ray source.
We examine scenarios for the 
spectra of CRHs
and/or primary electrons that
would reconcile 
all these different data. 
We determine that
(i) a hard ($\sim \, E^{-2.2}$), high-energy ($\gtrsim$ TeV) population CRHs
is unavoidably required by the HESS $\gamma$-ray data
and (ii) the remaining broad-band, non-thermal phenomenology
is explained either by
a rather steep ($\sim \, E^{-2.9}$) 
spectrum of primary electrons or
a ($\sim \, E^{-2.7}$) population of
CRHs.
Perhaps unsurprisingly, no single, power-law population of
either leptons or hadrons can explain the totality of broadband, non-thermal
Sgr B phenomenology.
\end{abstract}



\renewcommand{\thefootnote}{\arabic{footnote}}

\section{Introduction}

The HESS air Cerenkov telescope located in Namibia has recently made a ground-breaking detection 
\citep{Aharonian2006} of
a diffuse flux (over a solid angle of approximately 4$^\circ \, \times 2^\circ $)
of $\gamma$-rays in the 0.2-20 TeV (1 TeV = $10^{12}$ eV) range emanating from the region 
of the Galactic center
(GC) and distributed along the Galactic plane. This flux has been shown by the HESS collaboration
to be  correlated
with the density of molecular gas (mostly H$_2$) pervading the region, as determined from CS
measurements \citep{Tsuboi1999}, and argued, therefore, to  originate primarily from neutral
pion (and other meson) decay. These putative
mesons would themselves be generated in collisions between cosmic ray protons or heavier ions 
(generically, cosmic ray hadrons; CRHs) and
the ambient gas, the latter of which is contained in a number of giant molecular clouds complexes (GMCs), including 
the Sagittarius (Sgr) B complex. 
We label this broad idea `the hadronic scenario'.
Highly significantly for our purposes, the HESS collaboration has separately 
measured the $\gamma$-ray flux
emanating from a
$0.^\circ 5 \, \times 0.^\circ 5$ field centred on and covering
the Sgr B GMC itself (defined in Galactic co-ordinates by $0.^\circ 3<l<0.^\circ 8$ and $-0.^\circ 3<b<0.^\circ 2$) 
and the spectrum of this emission.

Within the hadronic scenario,
the well-known particle physics of pion (and heavier meson) production  and subsequent decay together with the HESS
$\gamma$-ray measurements allow us to make a prediction for the rate at which secondary electrons
and positrons (generically, secondary leptons) 
would be injected into the Sgr B cloud through the same CR collision mechanism that, by hypothesis, supplies the
$\gamma$-rays (CRH collisions lead to production of charged and neutral pions -- and charged and neutral
kaons -- in roughly equal numbers and 
the charged
pion and kaon decay chains ultimately terminate in the e$^\pm$ alluded to, together with neutrinos).
With the additional input of the local magnetic field strength, the gas density, and
the CRH spectral shape (viz., that the latter is a power-law
in momentum as generically predicted by shock acceleration theory with a \spin and \norm that match it smoothly
on to the HESS observations) 
we can predict the synchrotron radio spectrum from e$^\pm$ secondaries generated over the entire Sgr B GMC.
In fact, we can, on the basis of these inputs, 
self-consistently calculate the entire broadband spectrum of the Sgr B complex
from radio to TeV energies, accounting for all relevant radiative processes (secondary electron synchrotron, bremsstrahlung
and inverse Compton radiation as well as neutral meson decay).

The hadronic interpretation of the HESS data has recently been
challenged by \citet{Yusef-Zadeh2006} who, remarking on an independent correlation between Fe K$\alpha$
X-ray line emission and TeV $\gamma$-ray emission across the GC region, 
have suggested a model where a population (or populations) of {\it primary}
electrons is responsible for the non-thermal emission seen from a number of molecular clouds, including
Sgr B2. 
These authors suggest, in particular, that the TeV emission originate in inverse Compton (IC)
scattering of sub-millimeter
radiation from dust by the high energy component of the putative electron population. 
We label the broad idea that primary electrons are responsible for the bulk of the Sgr B phenomenology
`the leptonic scenario'.

We will examine the compatibility of both the hadronic and leptonic scenarios
with the broadband non-thermal spectrum of Sgr B and other Sgr B phenomenology
in detail in the current work.

\section{Inferred Environment and Morphology of Sgr B}

On the basis of  CS measurements \citep{Tsuboi1999}, the HESS collaboration report that the 
mass of molecular
material within the half degree by half degree field covering the Sgr B Complex is 
between 6 and 15 million solar masses. In the following 
we consider both these bounding values in modeling emission at various wavelengths from the region.

Sgr B is one of a number of GMCs bound
in relatively tight orbits (projected distance of $\sim$ 100 pc) around Sgr A*, 
the radio source associated with the supermassive ($\sim 3.7 \times 10^6 \msun$) 
black hole at the center of the Galaxy (at an assumed distance of 8.5 kpc).
Together these complexes constitute the central molecular zone, a structure that contains
fully  $\sim$10\% of the total molecular material of the Galaxy ($3 - 8 \times 10^7 \msun$
according to the CS measurements: \cite{Aharonian2006}, \cite{Tsuboi1999}).
The GC GMCs are characterised by supersonic internal velocity dispersions 
($\sim 15 - 50$ km s$^{-1}$: \cite{Morris1996}), 
high molecular gas kinetic temperature and increased metallicity
(\cite{Mayer-Hasselwander1998} and references therein).

The Sgr B complex contains two bright
sub-regions, Sgr B1 and Sgr B2, that show up in radio continuum observations
\citep{Mehringer1995,Yusef-Zadeh2006}. The latter is 
the largest molecular cloud in the Galaxy and contains one its largest
complexes of HII regions, being dominated by numerous ultra-compact and hyper-compact
H II regions which inhabit three dense cores (labeled North, Main, and South).
These are, themselves, located inside a structure labeled the Envelope \citep{Gordon1993}. 
There have been $\sim$57 UCHII
sources identified in Sgr B2(M) alone (Gaume \& Claussen (1990), Gaume,
et al. (1995) and De Pree, Goss, \& Gaume (1998)).  The cores are small
($\sim$0.5 pc), light ($10^{3-4}\solarmass$,
corresponding to $\sim$5\% of the Sgr B2 mass) and dense ($10^{6-
7}$ cm$^{-3}$).  On the other hand, the envelope is massive ($7.6\times10^{5}\solarmass$), and less dense
(10$^{5}$ cm$^{-3}$). The average $n_{H_2}$ across
Sgr B2 is $\sim 10^6$ cm$^{-3}$.

Sgr B1, a region of diffuse
thermal emission, is comparatively less studied than Sgr B2.  It is an HII region of
optical depth much less than one at $\sim$ cm wavelengths, suggesting that it is older, evolved structure
no longer containing the dense, hot
star forming UCHII regions found in Sgr B2
\citep{Mehringer1995}.

\subsection{Average Molecular Hydrogen Number Density of Sgr B}

In general, for molecular cloud material distributed through the CMZ at a distance $R_{\rm GC}$ from the 
GC, the minimal number density required in order that the cloud's self-gravity overcome 
tidal shearing by the central bar potential is given 
by \citep{Bania1986,Gusten1989,Stark2004,Morris2007}:
\be
n_{H_2}^{\rm min} \simeq 10^4 {\rm cm}^{-3} \left(\frac{75 {\rm pc}}{R_{\rm GC}}\right)^{1.8} \, .
\label{eqn_critDnsty}
\ee
The average\footnote{Note that this is the average over mass at a given density, 
not the volumetric average} molecular hydrogen number density over the entire
Sgr B complex $\left< n_{H_2} \right>$
is close to this critical value (given the $\sim$ 100 pc separation between Sgr B and the GC)
and lies in the 
$10^{(3.5-4)}$ cm$^{-3}$ range \citep{Lis1989,Lis1990,Paglione1998}. 
We adopt a working figure of $\left< n_{H_2} \right> = 10^4$ cm$^{-3}$ for definiteness.

As may be adjudged from the number densities quoted above, however,
one should keep in mind that the distribution of gas across Sgr B is highly clumpy.
Nevertheless, we have tested our models of synchrotron emission for injection into such a clumpy medium
(as parameterized on the basis of the results set out in Figure 6 of \cite{Paglione1998})
and find that integrated results are in agreement to within 20\% (small on the scale of
other uncertainties) with those obtained using a one-zone model 
-- as henceforth employed -- with
average values for molecular density (and magnetic field).

\subsection{Average Magnetic Field of Sgr B}
\label{section_BSgrB}

The determination of magnetic field strengths through the GC region
has been of long-standing interest; a very recent -- though not entirely disinterested review --
is that by \citet{Morris2007}.

The magnetic phenomenology of the GC is dominated by the so-called
non-thermal filaments (NTFs; \cite{Yusef-Zadeh1987,Morris1996}).
The NTFs are remarkable structures: they are 
magnetic flux tubes of up to 30 pc length but only fractions of pc wide, 
running predominantly perpendicular to the Galactic plane,
and illuminated by synchrotron emission. 
They are found at distances of up to 150 pc from the GC --
across a swathe of Galactic longitude, then, apparently conincident with the CMZ -- but not outside this region.
Given the almost invariant curvature of the filaments and their regularity -- despite clear interactions
with the turbulent GC ISM \citep{Morris2007} -- 
these structures must be quite rigid, consequently allowing the
inference of a very large field strength of $\sim$ mG scale \citep{Yusef-Zadeh1987,Morris1996}.

A long-standing and on-going controversy concerns the question of whether the $\sim$mG scale magnetic
field strength and structure determined for the NTFs is actually pervasive or not.
It is beyond the scope of this paper to consider this issue in any detail.
Instead, we shall adopt the approach of considering both the upper and lower scales for global GC magnetic field strength
argued for by various researchers in the following, viz., $\sim$ 1 mG and $\sim 10 \, \mu$G. 
The assumed strength of this global field shall be of importance when we come to consider 
whether a central object (located close to or at Sgr A*) is possible as the source for the populations of
relativistic particles we infer to be pervading Sgr B.

In any case, the poloidal
field structure traced by the NTFs
is observed (via far-IR polarization measurements: see \cite{Morris1996}, \cite{Novak2003}) 
to break down in GC molecular clouds. 
In fact, the field direction inside GC molecular clouds tends to be drawn out along a direction following the ridge
of molecular and thermal radio emission,
predominantly {\it parallel} with the Galactic plane, therefore and at right angles to the
NTF fields.
Such a  configuration might arise naturally from the shearing
action of the tidal field at the GC on the orbital motion on the molecular gas, 
drawing an initially poloidal configuration of field lines fixed in the matter into a toroidal one
as an inadvertant consequence of differential gas rotation (in a situation where
gravitational forces overcome magnetic ones).



As regards Sgr B in particular,
the average magnetic field of the complex, especially 
the average field magnitude over large scales, is
somewhat uncertain.
A determination of the line-of-sight magnetic
field strength $B_{\rm los}$ in the Sgr B2 region of the cloud
has also been made on the basis of 
Zeeman mapping the H I line in absorption
\citep{Crutcher1996}. This is  0.48 mG with a spatial variation of around 50 \%. 
Given the
resolution of those observations (5$\arcsec$ for Sgr B2), this
estimate may apply to the outer envelope of the cloud
complex \citep{Crutcher1996}. 
Further, the figure of $\sim$ 0.5 mG is actually a {\it lower limit}
to the average local field strength as line-of-sight field reversals
can only reduce the field strength inferred from Zeeman splitting and not increase it \citep{Novak1997}.
Also note that, statistically for a large ensemble of molecular clouds, $B_{\rm los} = 1/2 \, | {\bf B}|$
\citep{Crutcher1996}.

Sub-millimetre polarimetry has also been used to trace the topology of the Sgr B2 field 
\citep{Novak1997,Chuss2005} and, in concert
with the Zeeman measurement mentioned above, determines a lower limit to the large scale Sgr B2 field of 150 $\mu$G  
(to be distinguished from the local, presumably somewhat disordered, field directly measured as being at $\gtrsim$ 0.5 mG
by the Zeeman technique alone). 
The field structure of Sgr B is revealed by the most recent measurements to be quite complicated \citep{Chuss2005}.
The field of Sgr B2 in particular seems to follow a spiral, suggesting an ``extreme example of local forces
dominating the field structure" \citep{Chuss2005}.

Finally, using results obtained from an ensemble of 27 separate Zeeman
measurements of individual molecular clouds and assuming
uniform-density spherical clouds, \citet{Crutcher1999}
has derived the following scaling between magnetic field strength
and particle number density:
\be
B = 0.1 \left(\frac{n_{H_2}}{{10^4\mbox{ cm}^{-3}}}\right)^{0.47} \mbox{~~~mG}.
\label{eqn_Crutcher}
\ee
The Zeeman observation of Sgr B2 \citep{Crutcher1996} define
one of the data points from which this fit is obtained
(with an assumed $H_2$ number density of $10^{3.4}$ cm$^{-3}$), 
but it should be noted that the Sgr B2 data point actually falls furthest 
from (and considerably above) the fitted scaling relation (see \citet{Crutcher1999}, Figure 1). 
Nevertheless, 
the relation does capture the
approximate $\sqrt{n_{H_2}}$ scaling
for magnetic field strength
that
is predicted by many lines of theory (see \citet{Crutcher1999} and references therein), 
including
ambipolar-diffusion-driven star formation.
The scaling relation predicts, in our assumed average
$\left< n_{H_2} \right> = 10^4$ cm$^{-3}$, a field of 0.1 mG. 

\subsection{Hydrogen Column Density to and through Sgr B}

The hydrogen column density {\it to} Sgr B is estimated to be $\sim \, 1 \times 10^{23}$ H cm$^{-2}$ 
\citep{Murakami2001,Revnivtsev2004}. The column density {\it through} the densest part of the cloud,
the central 2' radius region of Sgr B2, is estimated to be $\gtrsim \, 8 \times 10^{23}$ H cm$^{-2}$
\citep{Lis1989,Murakami2001}. These two values bound the possible range for the average column density
to the emission in the one-zone models investigated in this paper (we actually fit to this parameter: see below).

\subsection{Temperature of Gas in Sgr B}

There is
evidence for a least two separate gas components across the CMZ:
a 150-200K component as traced by multi-transition ammonia lines (see \cite{Yusef-Zadeh2006} and references therein)
and a cooler and denser component with small volume filling factor. There is also
a 70 K dust component across the region.

Like the density structure, the temperature structure of Sgr B is complicated 
and it is therefore difficult to nominate a `typical' temperature for the gas in this complex.
In general across Sgr B2 there is a
strong temperature gradient ranging from about 100K
at higher Galactic latititudes and longitudes down to about 40K at lower latititudes and longitudes. 
The dense cores in Sgr B2 
are at higher temperatures, $\sim$ 300 K, due presumably to heating by young stars
and the surrounding, lower density envelope may be hotter still (600-700K: \cite{Wilson2006}).  
Furthermore, \citet{Wilson2006} have detected the (J,J) = (18,18) line of NH$_3$
in Sgr B2 which indicates a very hot component at 600-700K. 
Sgr B1 hovers at around 40-50K but does have some hot
spots of gas at temperatures up to 90K \citep{Ott2007}.

Also, as mentioned above,
the centimeter radio continuum maps show that Sgr B2 has a significant number 
($\sim$60) of compact, ultra compact (UC) and hyper compact (HC) HII regions, 
indicating extremely hot gas, at temperatures of few times $10^{3}$ K on size 
scale of 10's to 100's of AU \citet{dePree1998}, 
probably due to massive star formation in the dense cores of Sgr B2 Main and North.  

Keeping in mind that, given the clumpiness of the density distribution, much of the total mass within the complex
is located in the sub-structures named above, we estimate that the 
typical temperature for gas in the Sgr B complex lies in the range 150-200K.

\section{Radio Observations of Sgr B}

As described in detail below, 
we wish to compare the radio flux predictions we arrive at with
radio data we have collected from various sources.
For fluxes at 330 and 74 MHz ($\sim$90 cm and $\sim$400 cm) 
we make use of archival 
Very Large Array\footnote{The Very Large Array (VLA), as part of the National Radio Astronomy Observatory, is operated 
by Associated Universities, Inc., under cooperative agreement with the National Science Foundation.} 
data that covers the region in question \citep{Brogan2003}. 
At 843 MHz we make use of unpublished data obtained in the course of 
the Sydney University Molonglo Sky Survey \citep{Bock1999}\footnote{The Molonglo Observatory Synthesis Telescope (MOST)  
is operated by the University of Sydney}. 
Finally,
we have recently observed the dense cores of Sgr B at 1.384 and 2.368 GHz (13 and 20 cm) with the Australia
Telescope Compact Array\footnote{The Australia Telescope Compact Array (ATCA) 
is operated by the Australia Telescope National Facility, CSIRO, as a National Research Facility.} \citep{Jones2007}.
In the future we will perform observations with this array in a short baseline configuration that should allow the detection
of the non-thermal flux emanating from a relatively large solid angle we predict on the basis of the present work. For the 
moment, however, the data we have collected were obtained in a baseline configuration relatively insensitive to such flux
and provide, therefore, only a rather weak lower limit to the radio flux from the whole Sgr B complex at these wavelengths.
On the other hand, we have used these 13 and 20 cm measurement to probe the cosmic ray populations in a number of
sub-structures within Sgr B2: see \citet{Jones2007}.

We have obtained 74 and 330 MHz VLA maps \citep{Brogan2003}
of the Sgr B region both obtained with a beam of $\sim$1$\arcmin\times$2$\arcmin$.  
The
short baseline coverage with which these maps were obtained
implies a sensitivity to diffuse flux from scales of up to
6$\degree$ at 74 MHz and 1-2$\degree$ at 330 MHz. 

Fluxes at both 74 and 330 MHz were obtained by integrating within the
entire $0.5\degree\times0.5\degree$ region defined in \citet{Aharonian2006} 
and a smaller region of $0.35\degree\times0.30\degree$.
Unfortunately for our purposes,
74 MHz emission from the GC region is considerably
free-free absorbed by intervening HII regions \citep{Brogan2003,Kassim1990}
meaning that intrinsic flux is badly determined. 
Moreover, the Sgr B complex is not obviously detected
above background
in the map at 74 MHz so the total flux at this frequency
over the solid angle of the Sgr B complex only defines
a rough upper limit to the flux attributable to the object in question.

The determination of the 330 MHz flux was made by
taking the statistics
over the region, using the Miriad task \emph{histo}, and multiplying the
mean flux within the region by the number of beams within it.  
The flux derived is 45 Jy with an estimated (RMS) error for the primary beam corrected
image of 4 Jy. 

Following a similar procedure -- but noting the provisos outlined above --
the 74 MHz flux from the $0.5\degree\times0.5\degree$ region is 570 $\pm 120$ Jy.

\begin{figure}
\plotone{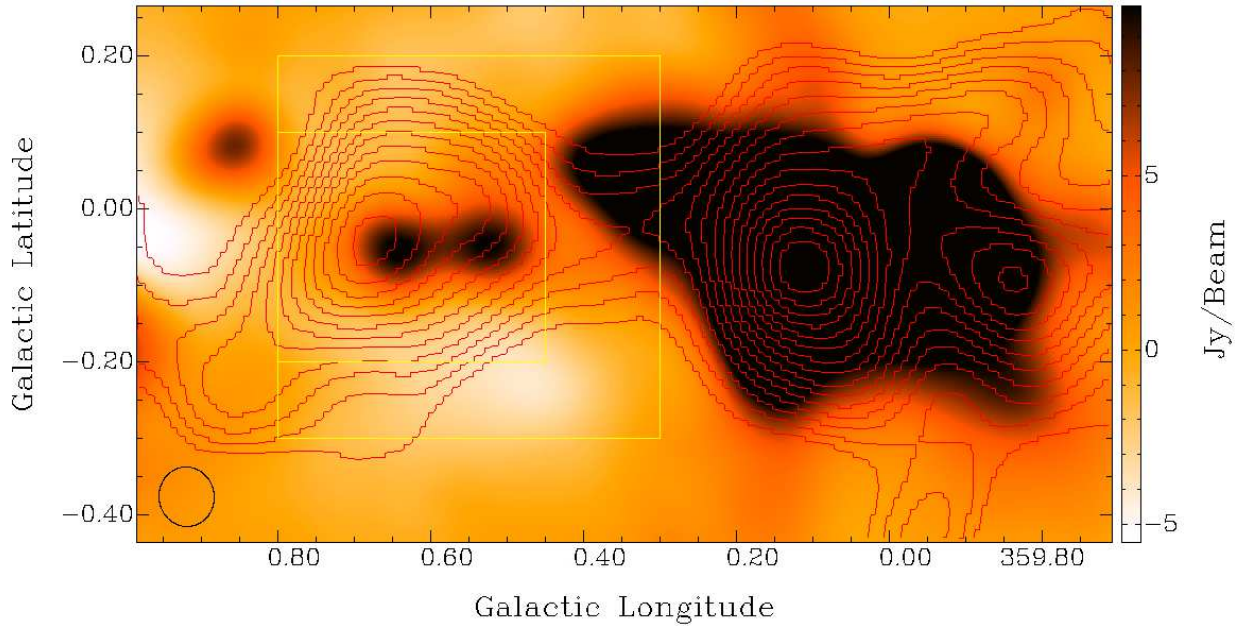}
\caption{Gray scale: 330 MHz VLA data 
with a 
position angle of 6$^\circ$ \citep{Brogan2003} smoothed to match 
the angular resolution of the HESS
instrument ($0.07^\circ$ = 252''; the circle shows the smoothed beam size). 
Contours: HESS ``foreground-subtracted" excess count map
for the GC region \citep{Aharonian2006} with $\gamma$-ray flux attributable to the  point-like
sources coincident with SNR G0.9+0.1 and Sgr A* removed. 
The contours are from 40\% to 90\% of total excess counts in increments
of 5\%. The larger (yellow) box indicates the $0^\circ.5 \times 0^\circ.5$ degree region
for which the HESS collaboration provides a $\gamma$-ray spectrum. We also calculate
the total radio and TeV $\gamma$-ray fluxes within the smaller region shown as described in the text.
The total radio flux within the smaller box accounts for more than half the total radio
flux within the larger box.}
\label{fig_90cm-smeared-HESS-boxes-overlay}
\end{figure}

We convolved the 330 MHz map  to the HESS resolution, using a Gaussian with a
standard deviation of 0.07$\degree \, = 252''$. 
This is shown as the gray scale of Figure \ref{fig_90cm-smeared-HESS-boxes-overlay}. 
The $0.5\degree\times0.5\degree$ field defined above and from which total radio fluxes
were calculated is shown as the larger of the two rectangles
and the circle is the smoothed beam size.
The peak at 330 MHz region is spatially correlated
with the peak of the HESS emission which is given by the contours and sits on top of the Sgr B2 structure.
The western lobe of the peanut-shaped radio emission feature sits on top of Sgr B1.

Using the above techniques, we also calculated the total 330 MHz flux from the smaller rectangle displayed in
Figure \ref{fig_90cm-smeared-HESS-boxes-overlay} which turns out to be 24 Jansky.
The ratio of radio to $\gamma$-ray fluxes agrees
to better than 20\% between the two regions, 
meaning that the correlation between radio and $\gamma$ emission is no better (actually slightly worse) for the 
smaller region when averaged over this scale. 
We will perform a more sophisticated analysis of the spatial correlation between
radio and $\gamma$-rays in a future work.

In the map one notes a clear radio signal from the SNR/pulsar wind nebula G0.9+0.1 which does not
seem to be correlated at all with $\gamma$-ray emission, but one should keep in mind here that this object {\it has} been detected
by HESS as a point source and its flux has been artificially removed from the $\gamma$-ray map as discussed below in
\S \ref{section_TeV}.

From unpublished data obtained in the course of the 
SUMSS data at 843 MHz we have also obtained
the total flux from the HESS-defined $0^\circ.5  \times 0^\circ.5$ field covering Sgr B.
Because of the strong sidelobes caused by the emission from Sgr A, however, we
can only make a crude flux estimate of 20 $\pm 10$ Jy. 


The spectral index between the central values of the 330 and 843 MHz fluxes is -0.9, with an allowable range
of -0.4 to -1.4 clearly indicating 
a non-thermal spectrum even granted the large errors on the 843 MHz flux. 

\section{Other Observations of Sgr B}

\subsection{X-ray Observations of Sgr B}

Galactic center molecular clouds, Sgr B2 in particular, have been intensively observed over the last decade in X-rays.
The X-ray data covering Sgr B2 reveal both strong continuum emission and a bright source of
fluorescent Fe K$\alpha$ line radiation within the cloud Sgr~B2
that displays an unusually large
equivalent width of $\approx 2$--$3$ keV.

In our data fitting we use the 
sub-set of the collection of broad-band X-ray data points assembled by \citet{Revnivtsev2004}
(and presented in their Figure 2; we re-display these points
in our Figures \ref{fig_plotH2aBB} 
and \ref{fig_plotL2eBB})
that approximately give only the {\it continuum} emission from Sgr B2. 
Note that these data points -- which are due to ASCA/GIS\citep{Koyama1996,Murakami2000}, 
GRANAT/ART-P, and INTEGRAL/IBIS over
ascending photon energy -- are for observations on somewhat smaller angular scales than the 
$0.^\circ 5 \, \times 0.^\circ 5$ field whose broad-band emission we are trying to model. 
They do, however,
cover a good fraction of the total mass in the Sgr B Complex given the clumpiness of the gas distribution
(around $2 \times 10^6 \, \msun$). 
In any case, because of the uncertainty introduced by this mismatch we do not try to reproduce
the continuum X-ray emission in minute detail 
in our modeling below. Instead we seek only to broadly match the normalization and
spectral index of this emission. 
Note also that {\it we do not seek to reproduce the Fe K$\alpha$ line emission at all in this paper}.

A cloud radiates via X-ray
fluorescence when it is illuminated, either internally or externally,
by a source of $\sim$ 8 keV X-rays or $\sim$ 30 keV CR ions or electrons. 
Now, a  steady X-ray source embedded within the
cloud produces an upper limit to the equivalent width of only $\sim 1$ keV
\cite[see, e.g.,][]{Fabian1977,Vainshtein1980,Fromerth2001}. 
A number of authors have taken the large equivalent width of Sgr~B2, then, as evidence for illumination by
an external X-ray source located towards the actual GC, usually identified with Sgr A*
(see \cite{Revnivtsev2004} and references therein).
In this scenario, then,
we are witnessing the X-ray echo, delayed by
300--400 years relative to the direct signal from the black hole, due
to the light travel time from the Galactic center out to Sgr~B2's
position (in which time,  by hypothesis, 
the direct source has fallen back into relative quiescence).
\citet{Fryer2006} have recently suggested that the large dynamic range 
(6 orders of magnitude: \citet{Revnivtsev2004,Fryer2006})
for Sgr A* required in this scenario is very difficult to account for theoretically and a more
plausible transient X-ray source of the appropriate magnitude and in the appropriate direction from Sgr B2
may have been provided by the collision of
the expanding shell of the Sgr~A East supernova remnant 
and the so-called 50 km s$^{-1}$ molecular cloud.

An alternative explanation to the hypothesised external X-ray illuminator has been advanced by
\citet{Valinia2000,Yusef-Zadeh2002} and \cite{Yusef-Zadeh2006}, viz.
that non-thermal electrons collide with ambient matter to produce inner-shell ionizations 
of iron atoms, leading to 6.4\,keV line emission, and
simultaneously generate bremsstrahlung radiation to supply the observed continuum X-ray emission.

Again the large equivalent width of the line emission is important here: 
the low energy electron  interpretation of the Fe K$\alpha$ emission 
would require a metallicity significantly higher -- a factor of 3-4 -- than solar
and  \citet{Revnivtsev2004} in fact raise this as a difficulty with the model. 
But it is already known
that GC MCs have metallicities 1.5-2 solar, in general. Thus as 
\citet{Yusef-Zadeh2006} remark, this scenario only requires that the metalicity of Sgr B2 be 1.5-2
that of surrounding clouds which, given the amount of star formation taking place in this object,
is perfectly credible.

In any case, because we do not actually attempt to reproduce the Fe K$\alpha$ line emission
from Sgr B2 in this paper, it is beyond the scope of this paper to go further into this controversy
(this issue will be addressed in a later work: \cite{Crocker2006}).
Rather, following the lead of \citet{Yusef-Zadeh2006}, in the modeling we present below, 
we will attempt only to reproduce (broadly) the continuum X-ray spectrum of Sgr B2 with bremsstrahlung emission from 
low energy electrons, either primary or secondary. 
Whether the Fe K$\alpha$ emission seen is due to an external X-ray illuminator
or, as in the model of \citet{Yusef-Zadeh2006}, is actually due to low-energy
cosmic ray electron collisions, is immaterial for our purposes: the models for
non-thermal particle populations we eventually arrive at would seem
to work in either situation.

\subsection{TeV $\gamma$-ray Observations of Sgr B}
\label{section_TeV}

Observations of the Galactic centre region by the HESS instrument shortly after its inception
led to 
the
detection of a point-like source of $\sim$TeV $\gamma$-rays at
the dynamical centre of the Galaxy 
(HESS J1745-2290: \cite{Aharonian2004}) and 
compatible with the position of the supermassive black hole
Sagittarius A* or the unusual supernova remnant (SNR) Sgr A East \citep{Crocker2005}. A deeper
observation of the region in 2004 revealed a second source: the SNR/pulsar wind nebula G0.9+0.1 \citep{Aharonian2005}.
Most recently, the HESS collaboration has demonstrated that, after
subtracting these two point sources from their map of the GC,
residual, fainter features are evident, in particular, 
emission extending along the Galactic plane for roughly 2$^\circ$ \citep{Aharonian2006}.
This foreground-subtracted map is displayed as the contours in Figure \ref{fig_90cm-smeared-HESS-boxes-overlay}.

The morphology of this diffuse emission correlates well with the
distribution of molecular material in clouds as traced by CS emission \citep{Tsuboi1999}
and the spectrum of this emission, which is detected by HESS over more than two orders of magnitude in photon energy,
is constant, within systematic errors,
across the entire region with a spectral index of $\sim$2.3. This is appreciably harder
than emission detected at $\sim$ GeV energies across the Galactic plane -- the origin of the latter being
roughly compatible with creation in collisions of a CRH population of the same shape as that observed locally, viz., 
a spectral index of $\sim$2.7.
Moreover, the HESS-detected $\gamma$-ray emission above 1 TeV is a factor of 3-9 times higher than in the Galactic
disk and would seem to require, therefore, a 
different or additional cosmic ray
population in this region. 

\subsection{GeV $\gamma$-ray Observations of Sgr B}

In contrast to its clear detection in $\gamma$-rays at $\sim$ TeV energies, Sgr B has not been detected in the 30 MeV - 30 GeV
energy range, despite being in the field of view of the EGRET telescope's lengthy observations of the GC region
\citep{Mayer-Hasselwander1998}. Indeed, it was explicitly noted by \citet{Mayer-Hasselwander1998} that
no localized excess associated with the Sgr B complex was detected by EGRET excluding the possibility of a significantly
enhanced CR density in these clouds -- {\it in the appropriate energy range}, of course.
The closest 
source EGRET did detect in these pointings, 3EG J1746-2851, 
was at first thought to have a localization 
marginally compatible
with Sgr A* or Sgr A East, but was later shown to be perceptibly off-set to Galactic east \citep{Hooper2005,Pohl2005}.
This imples that this $\sim$ GeV source can not be identified with the point source seen by HESS 
at/near Sgr A* (HESS J1745-2290:
\cite{Crocker2005}). The extent of the 3EG J1746-2851 source was closely investigated by \citet{Mayer-Hasselwander1998}.
They found that, though a point source could not be ruled out, a best fit single source model marginally implied
emission (above 1 GeV) such that a flux enclosure angle of radius 
0$^\circ$.6 was required to encompass 68 \% of the total flux. 
This is a solid angle larger than the largest under consideration here
(given by the
entire Sgr B TeV emission region of $0^\circ.5 \times 0^\circ.5$). Furthermore,
the fact that the 3EG J1746-2851 source was only marginally determined to be extended
(its HWHM was $0^\circ.7$ to be contrasted with the HWHM of the pulsar (GeV) point sources Vela, Geminga, and Crab at
$0^\circ.55$) means that at GeV+ energies the Sgr B complex would have been indistinguishable from a point source
were it detected by EGRET.

In concert, the non-detection by EGRET of Sgr B and its detection in the same pointings of 
the source 3EG J1746-2851 
imply an upper limit to the $\gamma$-ray emission from Sgr B of the same
level as the detected flux from 3EG J1746-2851 
(though it must be acknowledged here that putative Sgr B flux limits provided by 
low energy EGRET data points ($<300$ MeV) 
are more indicative than strict because of the growth in the EGRET psf towards lower energies). 
Moreover, these two, in concert with total molecular mass determinations for Sgr B,
further imply that
the density of CRHs through Sgr B at $\sim$ GeV-100 GeV energies must not be significantly higher than the local CRH density
in the same energy range as noted by \citet{Mayer-Hasselwander1998}. 
This is a significant constraint for our model fitting to obey as explained below. 

\subsection{Cosmic Ray Ionization Rate in Sgr B2}

Recent determinations \citep{vanderTak2006}
of the cosmic ray ionization rate $\zeta_{CR}$ in the Sgr B2 Envelope
from observations
of H$_3$ O$^+$ emission lines at 364 and 307 GHz with the Atacama Pathfinder Experiment (APEX) 
telescope\footnote{The
Atacama Pathfinder Experiment (APEX) 
telescope is operated by Onsala Space Observatory, Max Planck Institut f{\" u}r
Radioastronomie (MPIfR), and European Southern Observatory (ESO).} show that 
$\zeta_{CR}$ here is around 4 $\times 10^{-16}$ s$^{-1}$ (with a factor of four
uncertainty). The central value is thus an order of magnitude larger than
the value determined for local molecular clouds up to a few kpc from the Sun \citep{vanderTak2000}
and in the vicinity of the Solar System
on the basis of extrapolation of
measurements taken by the Pioneer and Voyager spacecraft, viz. 3 $\times 10^{-17}$ s$^{-1}$
\citep{Webber1998}. Van der Tak et~al. (2006) have also determined a 
lower limit for $\zeta_{CR}$ in Sgr B2(Main) of around 4 $\times 10^{-17}$ s$^{-1}$ and estimate
that the actual value of this quantity is $\sim$ 1 $\times 10^{-16}$ s$^{-1}$
The authors conclude that {\it the ionization rates of dense molecular clouds are mainly determined
by their location in the Galaxy through variations in the ambient CR flux} and that, as a second order
effect,  $\zeta_{CR}$ may be $\sim 3$ times lower in dense molecular clouds than in diffuse clouds.

In any case, these observations would seem to justify adopting an average   $\zeta_{CR}$ for the entire
Sgr B Complex of 4 $\times 10^{-16}$ s$^{-1}$ -- and this may actually be an underestimate.

\subsubsection{Cosmic Ray Heating in Sgr B}

In passing we note that, employing the results obtained by \citet{Suchkov1993} for the temperature of
molecular gas heated by cosmic rays and cooled by molecular line emission (see their Figure 1),
the cosmic ray ionization rate given above, together with the assumed average 
molecular hydrogen number density of $10^{4}$ cm$^{-3}$, translates into a temperature of $\sim$ 30 K.
It seems, then, that given the temperature range also stated above cosmic ray heating cannot
be the dominant heating mechanism in Sgr B.

\section{Secondary Particle Spectra from CRH Collisions}

We wish to model the injection of electrons, positrons and $\gamma$-rays into the Sgr B environment through the
decay of charged and neutral pions (and heavier mesons) which are themselves created in collisions between hadronic cosmic rays 
(protons and
heavier ions all the way up to Fe) and ambient gas (which is roughly 93\% H nuclei -- mostly in H$_2$ molecules in the
molecular cloud environment --
and 7\% He nuclei). To this end, we have employed two Monte Carlo event generators,  TARGET 2.1a \citep{Engel2003}
 and SIBYLL 2.1 \citep{Engel2000}, that simulate
proton-proton collisions at a given energy, to create numerical yield data for secondaries. 
We employed TARGET, in particular, to generate yields from single p-p collisions between 1.259 and 100 GeV
and SIBYLL to generate yields from 100 to $10^6$ GeV and these
two calculations were smoothed together at 100 GeV. 
We then used a MATHEMATICA routine to interpolate the yield data for energies intermediate to those directly
simulated.

Secondary spectra due to collisions between an arbitrary distribution of beam CR particles
with gas in the molecular cloud can be written as
\be
q_2 (E_2) = \int \, dE_p \, \frac{d N_p}{d E_p}(E_p) \, n_H \, \epsilon(E_p) \, \sigma_{pH}(E_p) \, \frac{d N_2}{d E_2}(E_p,E_2) \, ,
\ee
where $q_2$ is in secondaries/eV/s/cm$^{-3}$ with secondary $\in \{e^-, e^+, \gamma \}$;  
\be
d N_p(E_p)/d E_p = \frac{4\pi \, J_p(E_p)}{\beta(E_p) \, c} 
\ee
is the number density of cosmic ray p's per
unit energy pervading the medium; $J_p$ is the differential flux of cosmic rays protons at total energy $E_p$ in
cm$^{-2}$ s$^{-1}$ eV$^{-1}$ sr$^{-1}$; $n_H$ is the number density
of H nucleus targets in the gas; 
$\epsilon(E_p)$ is the (unitless) energy-dependent modification factor introduced
by \citet{Mori1997} to account for
the presence of heavier ions in both target and 
beam\footnote{This factor is determined from the observed terrestrial CRH spectrum and abundances.
It is constant with a value of $\sim$ 1.5 from CR proton energies 1 GeV - 100 GeV, climbing
slowly thereafter to be $\sim$ 1.85 at $10^5$ GeV.
This factor may, of course, actually 
be an underestimate in the case of the GC environment
where a higher supernova rate could imply more heavy ions in the CRH spectrum than detected locally
and a larger admixture of heavier nuclei in the target gas.}; 
$\sigma_{pH}(E_p)$ is the total, inelastic cross-section
for collisions between a beam proton at energy $E_2$ and a stationary
 target H nucleus (in cm$^{2}$ and as parameterized by \citet{Block2000}; and
$d N_2(E_p,E_2)/d E_2$ in units eV$^{-1}$ is the (differential) yield function obtained from the interpolation of the
Monte Carlos, i.e., the 
distribution of secondaries with respect to
secondary particle energy per interaction of a primary cosmic ray
p having energy $E_p$.

By way of reference, we have calculated the mean energy of the parent beam proton of a $\gamma$-ray
observed with energy $E_\gamma$ in a power-law distribution of given spectral index. We tabulate the results of this
calculation in an appendix. Similar tables are provided for secondary electrons and positrons.

\subsection{`Knock-on' Electron Production}

At low energies -- significantly below $10^8$ eV -- one must account for the fact that the primary source
of secondary leptons is no longer from meson decay but rather from thermal electrons directly `knocked-on'
by Coulombic collisions of primary protons and heavier CR ions. 
In calculating spectra of such electrons we use
the results set out in
\citet{Dogiel1990}.

\subsection{Relating $\gamma$-ray and Secondary Electron/Positron Production}

With the above technology we can find, for a given input CRH spectrum, the resulting emissivity of secondary particles.
In fact, we can tie together the emissivity of $\gamma$-rays and secondary leptons 
(due to both particle decay and the knock-on process) at a particular energy, 
given that we know the shape of the initiating CRH spectrum. In other words, we may write:
\be
q_e (E_e, {\rm spectrum}) = R_{e \gamma}(E_e,{\rm spectrum}) \, q_\gamma (E_e, {\rm spectrum})
\ee
where $q_e$ denotes the emissivity of secondary electrons + positrons and we have implicitly defined
\be
R_{e \gamma}(E_e,{\rm spectrum}) \equiv \frac{q_e (E_e, {\rm spectrum})}{q_\gamma (E_e, {\rm spectrum})}
\ee
in which the notation $q_2 (E_e, {\rm spectrum})$ indicates that the emissivity of the secondary 
is, in general, a function of the input spectrum of CRHs. For a given input spectrum shape, we can
numerically calculate $R_{e \gamma}(E_e,{\rm spectrum})$ using the results from our interpolation of the
pp collision Monte Carlos and, in addition, a contribution from the `knock-on' electrons identified above 
(note that in the calculation of $R_{e \gamma}$ $n_{H_2}$ factorizes out).
So, for instance, this quantity can be calculated for a
power-law spectrum of CRHs parameterized by spectral index $\gamma$ 
(the overall normalization of the CRH spectrum factorizes out):
\be
R_{e \gamma}(E_e,\gamma) = \frac{q_e (E_e, \gamma)}{q_\gamma (E_e, \gamma)} \, .
\ee
Even more generally, we can relate the emissivity of electrons + positrons at one energy to the 
$\gamma$-ray emissivity at another:
\be
q_e (E_e, {\rm spectrum}) = R_{e \gamma}(E_e,{\rm spectrum}) \, R_\gamma(E_e,E_\gamma,{\rm spectrum}) \,
q_\gamma (E_\gamma, {\rm spectrum})
\ee
where
\be
R_\gamma(E_e,E_\gamma,{\rm spectrum}) \equiv \frac{q_\gamma (E_e, {\rm spectrum})}{q_\gamma (E_\gamma, {\rm spectrum})} \, ,
\ee
which, again, we can of course particularize to the useful case of a power law spectrum of input CRHs.

\section{Steady-State Primary and Secondary Electron Distributions}

\begin{figure}
\plotone{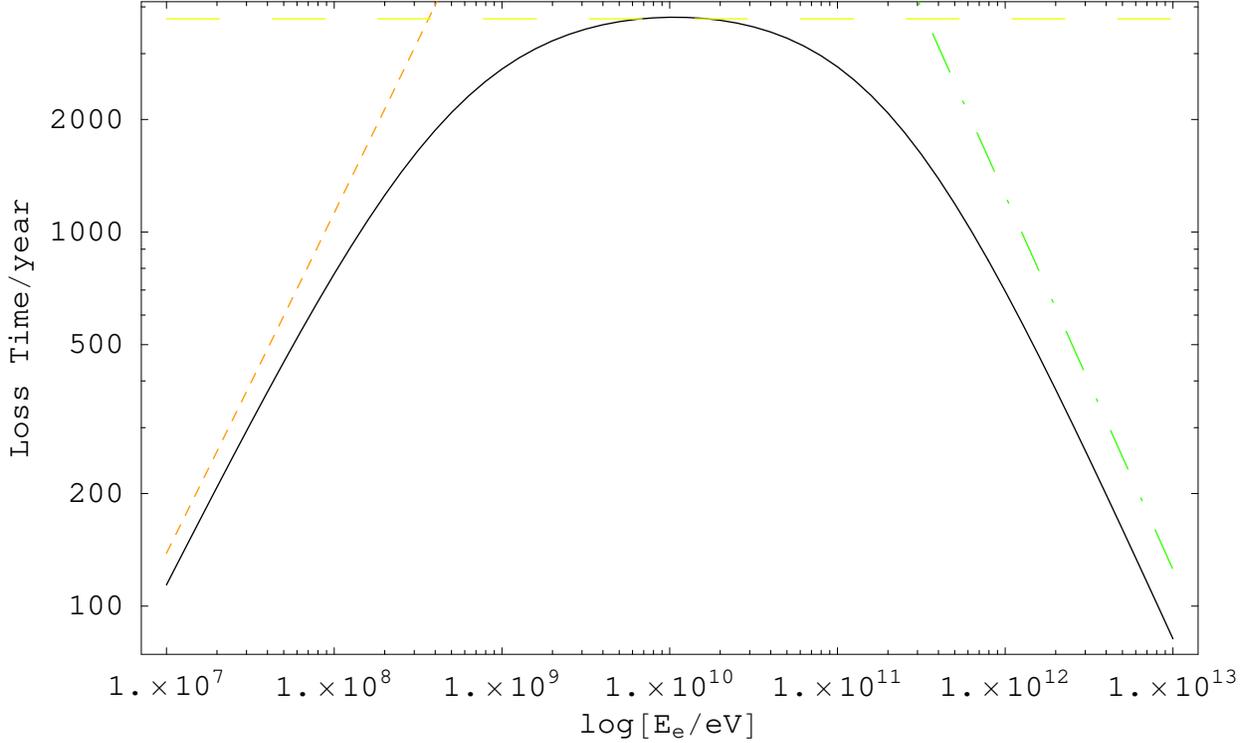}
\caption{Loss times for electrons and positrons in the Sgr B Complex average
environment ($n_H = 10^4$ cm$^{-3}$ and $B_{perp} = 10^4$ Gauss). The solid curve shows
the total loss time, the small dashed (orange) curve the ionization loss time, the 
long dashed (yellow) curve the bremsstrahlung loss time, and the long-short (green) dash curve shows
the synchrotron loss time. 
The IC losses are insignificant over this entire energy range
and always off the scale of this figure.
Note that a steady-state, loss-processed $e^\pm$ population
requires $\sim$4000 years to be established in the energy range $1\gtrsim E/{\rm GeV} \gtrsim 100$.
}
\label{fig_plotSgrBLossTimes}
\end{figure}

\begin{figure}
\plotone{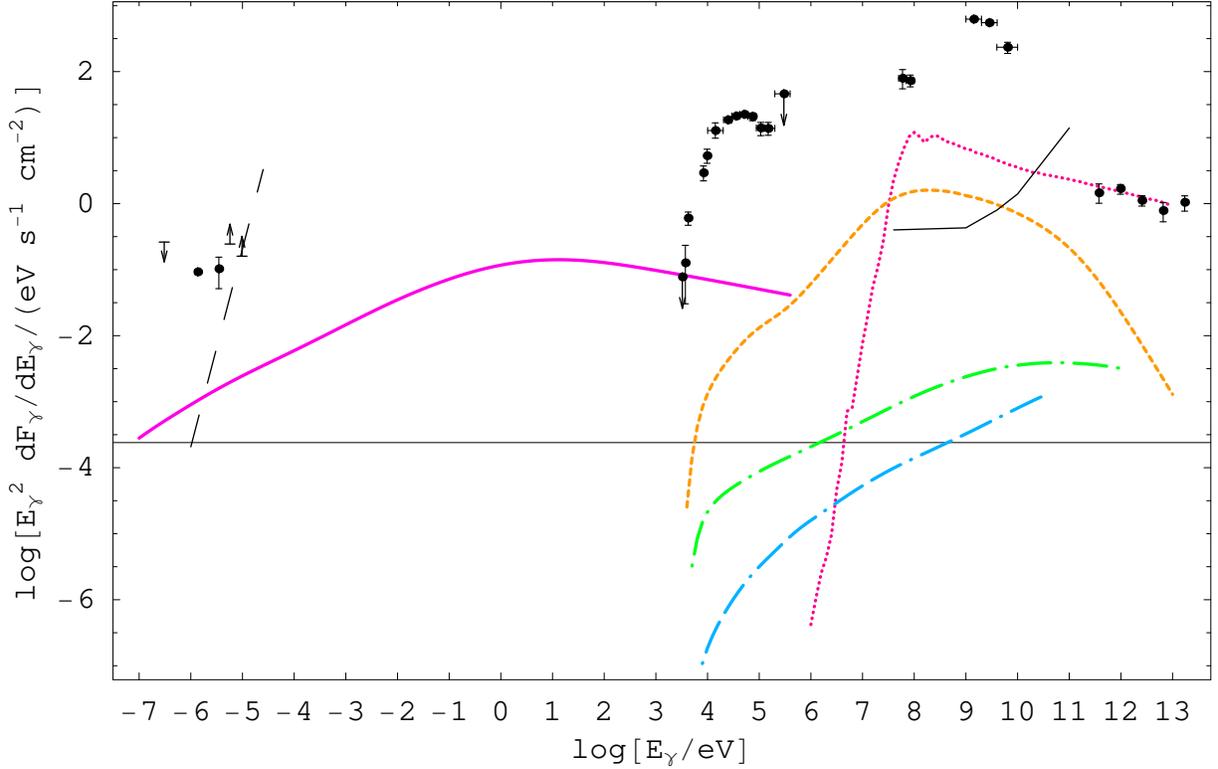}
\caption{\footnotesize The E$^2$-weighted broadband spectrum of the Sgr B complex
as predicted by a $\pi^0$-decay model for the TeV emission and assuming a 
pure power law in momentum for the colliding CRH's with a spectral index
of 2.2. A B field of 10$^{-4}$ as determined by the scaling
relation of \citet{Crutcher1999}, Eq.(\ref{eqn_Crutcher}), applied to the average density of the Sgr B Complex, viz. 
$\left< n_{H_2} \right> = 10^4$ cm$^{-3}$. 
{\bf Data}: radio data points at photon energies $\sim 10^{-(5-6)}$ are obtained from VLA, SUMSS, and ATCA
observations. The X-ray data points at $\sim 10^5$ eV are a sub-set of those
assembled by \citet{Revnivtsev2004} for the Sgr B2 cloud.
The EGRET data points at $\sim 10^8$ eV are a {\it de facto} upper limit
to emission from Sgr B at the indicated energy. The most constraining data points
from the spectrum of GC source \3EG have been selected here. The data points
at $\sim 10^{12}$ eV are from the HESS spectrum of Sgr B (the total $\gamma$-ray emission from the 
$0.^\circ 5 \times 0.^\circ 5$ field indicated by the larger rectangle in fig.(\ref{fig_90cm-smeared-HESS-boxes-overlay}).)
{\bf Fitted emission curves} -- 
dotted (red) line: $\pi^0$-decay $\gamma$-rays; 
solid (pink) line: synchrotron emission by secondaries;
short dash (yellow) line: bremsstrahlung emission by secondaries; 
dot-dash (green) line: IC scattered FIR light
(20 K, U$_\gamma$ = 5.7 eV cm$^{-3}$); 
dot-dot-dash (blue) line: IC scattered UV light
(30 000 K, U$_\gamma$ = 5.7 eV cm$^{-3}$); 
long dash (black) line: optically thick
thermal bremsstrahlung emission from the UCHII regions contained in Sgr B(Main) and
Sgr B(North) with an assumed temperature of 10 000 K and solid angles as given by \citet{Gordon1993}.
Photo-absorption (at X-ray energies) in the assumed $8 \times 10^{23}$ cm$^{-2}$ column density
is {\it not} accounted for in calculating the theoretical synchrotron curve.
The primary hadrons initiating the secondary leptons that produce the sycnhrotron are assumed
to cut off sharply at $10^{16}$ eV leading to a mirroring cut-off in the synchrotron spectrum at
$\sim 4 \times 10^5$ eV.
As is immediately apparent, 
the model fails to explain any data apart from 
the gamma-ray flux.
{\bf Sensitivity curve:} the solid (concave up, black) curve shows $E_\gamma \times$ the projected
integral flux
sensitivity of the GLAST \citep{Carson2006} instrument for a 5 $\sigma$ 
detection of a point source in 1 year and assuming a $\propto E_\gamma^2$
spectrum: see the GLAST webpage, http://www-glast.stanford.edu.}
\label{fig_plotH2aBB}
\end{figure}

\begin{figure}
\plotone{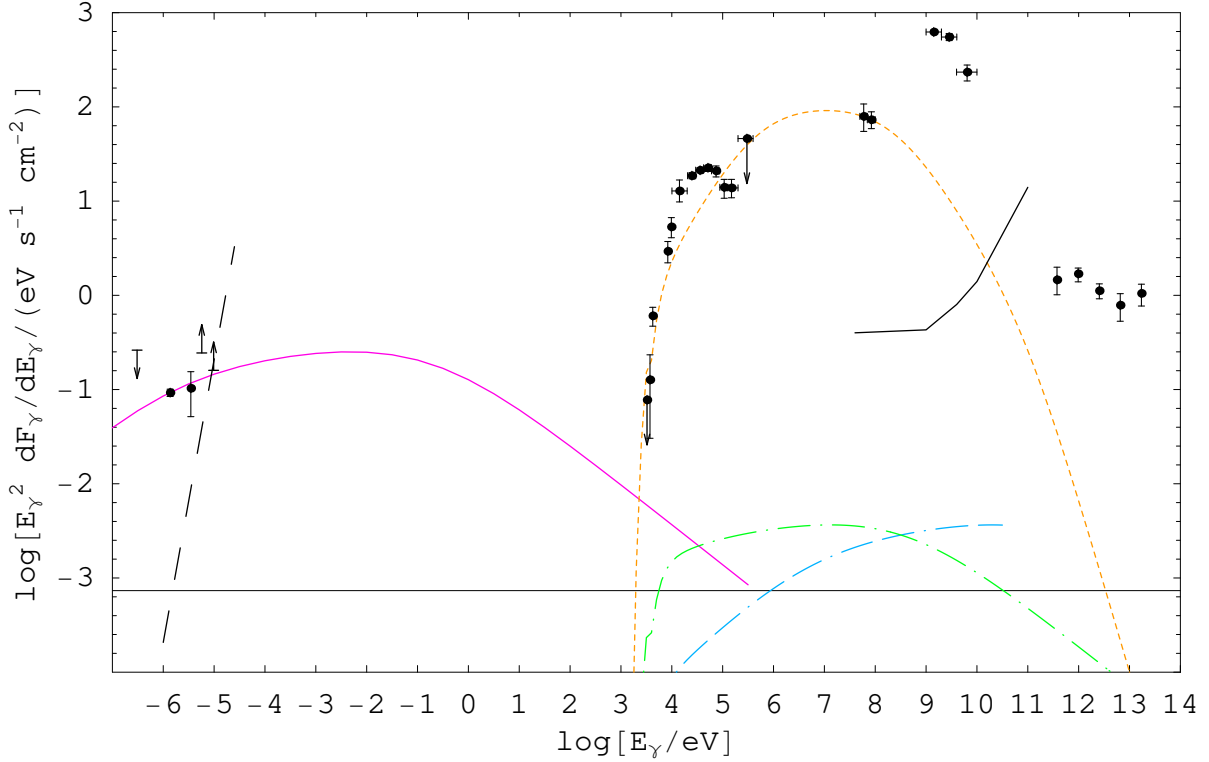}
\caption{The E$^2$-weighted broadband spectrum of the Sgr B complex
as predicted by the primary electron model set out in the text:
the primary electron normaliztion and spectral index are determined by
fitting to the ionization rate and, broadly, to the X-ray spectrum
(we do not seek to reproduce this exactly as described in the text). 
The B field (=1.3 $10^{-4}$ G)
is that required for this spectrum to reproduce the 90 cm datum and the assumed, average
molecular hydrogen number density is 
$5 \times 10^3$ cm$^{-3}$ (in order that bremsstrahlung emission obey the EGRET limit).
Photo-absorption in the fitted $4 \times 10^{23}$ cm$^{-2}$ column density
is {\it not} accounted for in calculating the theoretical synchrotron curve.
The synchrotron curve is taken to cut off sharply at $\sim 4 \times 10^5$ eV 
mirroring an assumed cut-off in the primary electron spectrum at $\sim 10^{14}$ eV.
Processes
corresponding to various curves are as described in fig. \ref{fig_plotH2aBB}.
}
\label{fig_plotL2eBB}
\end{figure}

For the Sgr B environment, we have $n_H \sim 10^4$ cm$^{-3}$ and average
magnetic fields $B \sim 10^{-4}$ Gauss, implying relatively short energy loss
times for electrons and positrons: 
e.g., the loss time $t_{\rm loss} (E_e)$ for leptons 
in this environment reaches a {\it maximum} of 
$\sim 3500$ years for energies around 10 GeV (see Figure \ref{fig_plotSgrBLossTimes}).

Note that in this work we do not consider non-steady-state lepton distributions at significant length. 
We introduce this restriction basically to avoid opening up the parameter space
by too much: non-steady state models -- which will necessarily involve
fine-tuning given the short loss timescales mentioned above (see \S \ref{sectn_dblpwrlw}) -- 
will be investigated in detail in a later
work \citep{Crocker2006}. 

Now, noting that (i) synchrotron radiation by electrons of $\nu$ GHz frequency
will be generated by
electrons of energy 
$E_e^{\rm corres} \simeq {\rm GeV} \sqrt{(\nu/{\rm GHz}) \, (10^{-4} \, {\rm Gauss}/B)}$
in the expected magnetic field strength,
(ii) the loss time at GeV is $\sim 3000$ years,
and (iii) taking an absolute upper limit
on the diffusion co-efficient in the molecular cloud environment 
to be given by the value appropriate to the Galactic plane, viz.
$D(E) \sim 5.2 \times 10^{28} \, (E/3 \, {\rm GeV})^{0.34}$ cm$^{-2}$ s$^{-1}$
\citep{Ptuskin2006}, an upper limit on
the diffusive transport scale is given by
$\sqrt{2 D({\rm GeV}) t_{\rm loss} ({\rm GeV})} \sim 20$ pc, approximately equal to
the radius of the Sgr B Complex.
This is likely to be a considerable overestimate of the diffusive transport scale
given that, in the turbulent magnetic environment of
GC molecular clouds, the diffusion coefficient is likely to be considerably
suppressed with respect to its value in the Galactic plane.
The above means that diffusive transport
can be neglected and, therefore, the ambient number density
of electrons + positrons, per unit energy, 
$d n_e(E_e, \vec{r})/d E_e$
(in $e^\pm$ cm$^{-3}$ eV$^{-1}$),  at various positions $\vec{r}$ within
the molecular cloud complex, can be obtained in steady-state
by numerical integration :
\be
\frac{d n_e}{d E_e}(E_e, \vec{r}) = 
{\int_{E_e}^\infty q_e(E_e', \vec{r}) dE_e' \over - dE_e(E_e)/dt} 
\label{eqn_prcsdSpctrm}
\ee
where $q_e(E_e', \vec{r})$ is the injected emissivity
of electrons + positrons that, in general, might
be either primary (i.e., directly accelerated {\it in situ}) or secondary and
$dE/dt(E_e)$ is the total rate of energy loss of electrons at energy $E_e$ due to ionization, 
bremmstrahlung, synchrotron and, possibly, IC emission (because of the energies involved, we neglect positron annihilation, 
and assume electrons and positrons suffer identical energy losses).  
Electrons and positrons lose energy by ionization losses in neutral molecular hydrogen at a rate
\be
{dE_e\over dt}_{\rm ioniz}\simeq  - \frac{1.5 \times 10^{-8}}{\beta_e} \times \left({n_{H_2}\over {\rm cm^{-3}}}\right) \times
\ln\left[\frac{E_e^2 \, \beta_e^2 \, (\gamma_e - 1)}{2 \, E_{exctn}^2}\right]\mbox{ ~~ eV/s} \, ,
\ee
where $E_{exctn}$ is the average excitation of the medium 
(in the molecular cloud environment of interest we set $E_{exctn} = 15$ eV -- see \cite{Schlickeiser2002} p. 99).
Electrons and positrons lose energy by bremsstrahlung in molecular hydrogen at a rate
\be
{dE_e \over dt}_{\rm bremss} = - 1.7 \times 10^{-15} \times \left({n_{H_2}\over {\rm cm^{-3}}}\right) \times \left(\frac{E_e}{\rm eV}\right) \mbox{ ~~ eV/s} \, .
\ee
The synchrotron energy loss rate is
\be
{dE_e \over dt}_{\rm synch}= - 1.0\times 10^{-3} \times \left({B_\perp \over {\rm 1 \; Gauss}}\right)^2 \times  
\left(\gamma_e^2 \, -1 \right) \mbox{ ~~ eV/s},
\ee 
where $B_\perp$ is the component of magnetic field perpendicular to the electron's direction.  
For an isotropic electron population $\langle B_\perp \rangle=0.78 B$.

\subsection{Cooled Primary Distribution}

In the case that $q_e$ is attributable to a steady-state injection of {\it primary} electrons, we have no independent, empirical
handle on this {\it injected} population -- we only see emission from the {\it cooled} distribution of electrons.
This is to be contrasted with the situation for secondary leptons where high-energy $\gamma$-ray emission 
from neutral meson decay can always provide, in principle, such a handle as explained above.

The cooled electron distribution
is steepened at high energies
by synchrotron radiation and flattened at low energies by ionization losses. 
However, for an intermediate range of energies 
around $\sim$ GeV
(depending on the exact details of the magnetic field and ambient gas density) -- roughly the same energy
range as electrons that are synchrotron radiating at $\sim$ GHz wavelengths -- 
bremsstrahlung emission is the dominant cooling process
and, because the bremsstrahlung cooling rate
has a linear dependence on electron energy, it does not modify an injection spectrum. 
In particular,
cooling of a power-law distribution of electrons injected with a spectral index $\gamma$
will not modify this spectral index around the $\sim$ GeV scale in a typical molecular cloud environment.
These two facts, together
with the relation between  spectral index of the synchrotron-radiating
electron (power-law) distribution and the spectral index of the generated radio spectrum given below
in Eq.(\ref{eqn_SPIN}),
imply that 
spectral index measurements between radio fluxes taken in $\sim$ GHz range
may give a handle on the {\it injected} electron distribution spectral index. 
Furthermore,
once this fixed normalization point has been obtained, provided the magnetic and 
gas environment of the region in which the synchrotron emitting electrons is located is known,
the spectral distortions introduced by ionization and synchrotron, at low and high energies respectively, 
may be calculated
so that the overall shape of the cooled electron spectrum may be determined. 
Finally, the overall normalization of the cooled
distribution
may also be obtained given that the distance to the emitting region is known.
Putting all these together -- as we do in a numerical routine -- and assuming {\it a priori}
that the injection spectrum for primary electrons should be a power-law in momentum (as described above) 
allows one to obtain the injection spectrum: this is simply the power-law
distribution characterised by an overall normalization and spectral index that, once cooled by the ionization, bremsstrahlung, 
and synchrotron processes in the known $n_{H_2}$ and magnetic field reproduces the observed radio spectral index and overall flux.

\section{Radiative Processes of Interest}

\subsection{Synchrotron Emission}

The synchrotron emission coefficient (for both primary and secondary electrons) can be calculated using standard
formulae in synchrotron
radiation theory \citep{Rybicki1979}
\begin{eqnarray*}
j_\nu(\nu,\vec{r})&=&  1.87 \times 10^{-30} \,\frac{\rm Watt}{{\rm Hz \, sr \, cm}^3} \, \times
\left({B_\perp\over {\rm \; 1 \; gauss}}\right)\int_{m_ec^2}^\infty F(\nu/\nu_c)\, 
\frac{d n_e(E_e, \vec{r})/d E_e}{\rm cm^{-3}} \, dE  \, \,  , \\
\nu_c &=&  4.19\times 10^6 \times (E / m_ec^2)^2 \times (B_\perp/{\rm \; 1 \; gauss}) \, \, {\rm Hz},\\
F(x) &=& x\int_x^\infty K_{{5 \over 3}}(\xi)d\xi,
\end{eqnarray*}
where 
$K_{{5 \over 3}}(x)$ is the modified Bessel function of order 5/3.

Note that, to a good approximation, for a synchrotron-radiating, power-law distribution of electrons with
spectral index $\alpha$, the observed radio spectrum will also be governed by a power law of index $\alpha$ where
\be
\alpha = \frac{\gamma - 1}{2} \, . 
\label{eqn_SPIN}
\ee

\subsection{Bremsstrahlung and Inverse Compton Emission}

The differential power density originating from bremsstrahlung collisions of $e^\pm$ pervading a medium with atomic number $Z$ and
number density $n_Z$ can be determined to be
\be
\frac{d P_\gamma^{\rm brems} }{d E_\gamma}(E_\gamma,\vec{r}) = 1.7 \times 10^{-15} {\rm \frac{eV}{s \, cm^3 \, eV}} 
\times 
\int_{m_e c^2 + E_\gamma}^\infty \, \frac{d n_e(E_e, \vec{r})/d E_e}{\rm cm^{-3}} \, d E_e \, .
\ee

Similarly, IC emission by a population
of electrons $d N_e(E_e)/d E_e$ off a target, thermal photon field characterised by temperature $T$ is approximately
given by
\be
\frac{d P_\gamma^{\rm IC} }{d E_\gamma}(E_\gamma) \simeq 1.36 \times 10^{-5} \, {\rm s}^{-1}
\times 
\left(\frac{T}{\rm K}\right)^{5/2} 
\times
\left(\frac{E_\gamma}{\rm eV}\right)^{-1/2}  
\times
\frac{d N_e}{d E_e}\left(E_e = 3.35 \times 10^7 \, {\rm eV} \, \sqrt{ E_\gamma/{\rm eV} \times {\rm K}/T}\right) \, ,
\ee
where the Thomson regime is assumed requiring
$\sqrt{E_\gamma/{\rm eV} \times T/{\rm K}} < m_e/\sqrt{2.7 \, k_B \, {\rm K \, eV}} \simeq 3.35 \times 10^7$. 
For the Sgr B photon background we assume -- to generate an upper limit --
energy densities in ultra-violet and
dust-reprocessed infra-red light fields
the same as that determined for Sgr A East SNR at much smaller Galactic radii ($\sim$ 10 pc), 
viz., both 5.7 eV cm$^{-3}$ \citep{Melia1998}.
The temperatures of these
distributions are taken to be at 30 000 and 20 K respectively. 
Despite this over-generous attribution of energy into these light fields we find that 
IC cooling and emission are always sub-dominant to other cooling and radiative processes
across the entire lepton spectrum (see below).

\subsection{Relating Synchrotron, Bremsstrahlung, and Inverse Compton Emission to $\gamma$-ray Emission}

With the technology outlined above we can predict on the basis of an observed flux of $\gamma$-rays 
from a particular astrophysical object or region -- 
which must, by hypothesis, orgininate in
hadronic interactions of a primary CRH population (i.e, mostly from neutral pion decay) -- the 
radio flux from the same object/region due to secondary leptons. 
Apart from the $\gamma$-ray flux, the inputs to the synchrotron prediction are 
(i) the
{\bf B} field strength in the object/region, (ii) the number density of target particles,
in our case,
molecular hydrogen, $n_{H_2}$, and (iii) the 
shape of the cosmic ray spectrum between 
(a) the energies of CRHs which are responsible for the observed $\gamma$-rays 
and
(b) the energies of CRHs which are responsible for
the secondary e$^\pm$'s whose synchrotron emission we might detect in the radio in some given frequency range.

Significantly, because the synchrotron expectation is normalized directly to the $\gamma$-rays,  a
determination of either the total mass of target particles 
or the distance to the object/region containing the targets
is not necessary in this calculation.

We  can use very similar considerations as above to determine predictions for bremsstrahlung emission by
secondary leptons. Likewise, with the additional input of the target photon field (radio, IR, optical, UV)
we can predict the IC emission by this population of secondaries.

Putting all this together, we can assemble the predicted broad-band spectrum of a particular astrophysical
object or region. The calculation is only self-consistent, of course, 
if, at the $\gamma$-ray energy where we normalize our predictions,
the $\gamma$-rays do strictly originate only in CRH collisions, though a small level of `pollution' due to
IC or bremsstrahlung at the $\lesssim 20\%$ level is perfectly acceptable given other uncertainties.

\section{Predicted Sgr B Phenomenology in Hadronic Scenario}

\subsection{Predicted Sgr B Broad-band Spectrum in Hadronic Scenario}

As explained above, adopting the expected, average values {\bf B} $= 10^{-4}$ Gauss and $n_{H_2} = 10^4$ cm$^{-3}$ and assuming a 
power law in momentum for the parent CRH population that, by hypothesis, is responsible for initiating the $\sim$ TeV $\gamma$-ray 
emission (with spectral index $\simeq 2.3$ and a normalization consistent with this emission), we may determine the broadband 
spectrum of Sgr B due to all pertinent electromagnetic emission mechanisms (initiated by primary CRHs and secondary leptons).
This spectrum is displayed in Figure \ref{fig_plotH2aBB}. Here one can see immediately that
the model fails to explain any data apart from 
the gamma-ray flux -- though it is not, of course, excluded by the other data. 
Of particular note, secondary synchrotron does not account for the radio flux detected at 330 MHz 
(radio datum second from left) or 843 MHz.

One might seek to `dial-up' the magnetic field strength in the model in order to account for the radio data, but in this
endeavor two problems are encountered: (i) the spectral index of the predicted spectrum is rather too flat to go through 
both 330 and 843 MHz data points (the spectral index between predicted central flux values is -0.31, cf.
the observational value of -(0.9 $\pm$ 0.5)) , even allowing for 1-$\sigma$ excursions of both experimental values and, more tellingly,
the magnetic
field strength demanded is unreasonably high at 10 mG (as supposed average over the entire Sgr B Complex), implying
a magnetic field energy density of $2.6 \times 10^6$ eV/cm$^{-3}$.

\subsection{Cosmic Ray Ionization Rate in Hadronic Scenario}

Even if we do not rule out such a field {\it a priori}, there is another reason to reject this model:
the contribution of the primary CRH population inferred from the $\gamma$-rays is far too small to account for the observed
cosmic ray ioniztion rate $\zeta_{CR}$ in Sgr B2, viz., 4 $\times 10^{-16}$ s$^{-1}$ as given above. And
as stated already, this value
should roughly apply over much of the Sgr B Complex. 

We may calculate the predicted CRH ionization rate for Sgr B using the technology set out in
\citet{Webber1998}.
Employing the Bethe cross-section 
\be
\sigma_{\rm Bethe} (\beta, Z) = 1.23 \times 10^{-20} {\rm cm}^2 \frac{Z^2}{\beta^2} 
\left[6.2 + \ln \left(\frac{\beta^2}{1 - \beta^2}\right) - 0.43 \, \beta^2 \right]
\ee
the ionization rate is
\be
\zeta_{CR}(J_p, m_{CR}) =  \frac{5}{3}  \,
    \int_{T_p^{\rm min}} ^{\infty} \, 4 \pi \, J_p(T_p) \, \sigma_{\rm Bethe} \, d T_p \, ,
\label{eqn_InztnRt}
\ee 
where $J_p$ is the differential flux of cosmic rays protons at kinetic energy $T_p$ in
cm$^{-2}$ s$^{-1}$ eV$^{-1}$ sr$^{-1}$ and we take $T_p^{\rm min} \equiv $ 10 MeV. 
To account for the effect of heavier ions in the CRH flux, one may introduce a constant multiplicative factor on the RHS of
equation (\ref{eqn_InztnRt}) which is 1.89 for the local CR flux \citep{Spitzer1968}
We note in passing that the above can be directly adopted for calculation of ionization by electrons, though in this case
the appropriate energy at which to cut-off the ionization rate integral is $T_e^{\rm min} \equiv $ 3 MeV \citep{Webber1998}.

Using these results, the predicted
$\zeta_{CR}$, is only 6 $\times 10^{-18}$ s$^{-1}$ and, if the entire non-thermal radio flux is accounted for by secondaries
generated by this CRH population, there is no room left for another low-energy particle population whose collisions
might account for the $\zeta_{CR}$ measurement. 
Even allowing for a possible additional contribution from heavier ions in the GC
environment (relative to their contribution in the Solar System environment) could not make up the 
$\sim$ two order of magnitude deficit
with respect to the $\zeta_{CR}$ measured by \citet{vanderTak2006} in Sgr~B2.

As a further confirmation that the TeV $\gamma$-ray and ionization data are irreconcilable within the current CRH model
one may turn around the above logic: for a power-law distribution of CRHs with a normalization
sufficient to account for the observed $\zeta_{CR}$ one may determine
what $\sim$TeV $\gamma$-ray spectrum is implied. Carrying out this exercise one finds that the 
$\gamma$-ray flux from Sgr B is overpredicted by around two orders of magnitude.

Of course, it is to be admitted here that to posit a meaningful relation between the ultra-relativistic
cosmic rays that generate the TeV $\gamma$-rays and the sub-relativistic particle population primarily responsible 
for ionizing the ambient gas might be considered speculative. 
Nevertheless, the result stands that these two phenomenological inputs are irreconcilable with the simplest
expectation from shock acceleration theory, viz., that the ambient CRHs be governed by a single power law in momentum.

\subsection{Double Power-Law Hadronic Scenario}
\label{sectn_dblpwrlw}

One simple and reasonable modification of the above scenario is to allow for two power-law CRH populations: a steep
population, dominant at low energy and with a normalization given
by the requirement that it reproduce the observed ionization rate 
and with a spectral index as flat as possible given the
EGRET constraint (implying a value of at least 2.7) and a flatter population of spectral index
of $\sim$ 2.2 that becomes dominant at high energy and explains the TeV $\gamma$-ray emission. 
Note that in the GC region it is not at all unreasonable
that such a complicated overall spectrum might pertain:
the region posseses two and possibly
three sources of cosmic rays --
unlike other regions in the Galaxy there
is activity here associated with the super massive black hole, there is a
very high supernova rate and the GC
molecular clouds represent a huge reservoir of energy in their turbulent motions.  

In any case,
we have investigated a double power law
scenario and indeed find that for a rather strong average magnetic field, viz., 2.2-3.7 mG (the range implied
by the range for the possible total mass of Sgr B Complex),
the broadband non-thermal phenomenology is well reproduced.
The required energy density in this assumed population is $\sim 2.3$ eV cm$^{-3}$, very similar to the local CR
energy density \citep{Webber1998} and, interestingly, the spectral index of this inferred population 
is very similar to the local one (though one should note that the inferred spectrum, $\propto p^{-2.7}$,
is substantially in excess of the local CRH spectrum at 10-100 MeV, explaining how it is possible the former can
produce ten times the ionization rate of the local spectrum, despite possessing approximately the same total
energy). We have also checked the steady-state positron production rate in this scenario and find that
the scenario predicts a 511 keV $\gamma$-ray production rate from 
electron-positron annihilation of $\sim 3 \times 10^{46}$ yr $^{-1}$, well
inside the limit by INTEGRAL observations ($\sim 10^{50}$ yr $^{-1}$ out to an angular radius of 8$^\circ$: 
\cite{Knodlseder2003,Jean2003}).
The question of the naturalness of this scenario is addressed in the Discussion section but one should note
immediately that the energy density in the required magnetic field
is 400-1600 times larger than that in the ``expected" field given by the Crutcher scaling.
In a future publication \citep{Crocker2006} we will also investigate 
an inflected primary CRH spectrum -- asymptoting to flat power laws at high and low energies but rather steeper at
energies in the GeV - 100 GeV range -- that holds out the promise of reproducing the broadband phenomenology
but for a rather smaller magnetic field.

\subsection{Non Steady State Scenario}

As stated, we have assumed a steady-state secondary lepton distribution in our calculations above requiring implicitly
that the age of the injected lepton distribution is greater than the cooling timescale.
If this condition is not satisfied, 
then, in place of Eq.(\ref{eqn_prcsdSpctrm}) one has, to tolerable accuracy \citep{Fatuzzo2005}, that
the secondary distribution is given by (injection spectrum) $\times$ (age of lepton population).
Now, in consultation with Figure \ref{fig_plotSgrBLossTimes}, one also notes that the 
loss time is a strong function of energy with different loss processes dominant at different 
enegry scales. 
In fact, one can see from this Figure that it would be possible to fine-tune the assumed age
of the lepton distribution in such a way that the leptons predominantly responsible for the
observed radio emission (i.e., those at $\sim$ GeV) be out-of-equilibrium, whereas those at lower and
higher energies would be in steady state.
Now, although the absolute emissivity of the lepton distribution reaches its greatest value when steady state is reached,
the greatest emissivity of GeV-scale leptons {\it relative} to the $\gtrsim$ TeV energy $\gamma$-ray emission
would be achieved in such a 
fine-tuned situation (that would require a lepton population of age $<3000$ years).
Thus, a larger $\sim$ GHz radio emission could be achived for the same observed $\sim$ TeV $\gamma$-ray
emission in this somewhat fine-tuned scenario relative to a completely steady state model assumed above.
This scenario will be investigated at greater length in a forthcoming publication \citep{Crocker2006}.

\section{Predicted Sgr B Phenomenology in Leptonic Scenario}

The above considerations naturally lead one to consider the possibility that primary electrons, 
presumably accelerated in-situ given the short loss times they experience in 
this strong magnetic field and dense environment, can be invoked to explain the Sgr B phenomenology
under consideration.

\subsection{The Model of Yusef-Zadeh et al.}

Recently \citet{Yusef-Zadeh2006},
motivated
by their observation of a three-way correlation
between the distribution of molecular material across the GC region, Fe K$\alpha$ line
emission, and TeV emission, have considered a model in which a power-law population of primary
electrons
is invoked to explain the keV to TeV emission from a number of dense molecular regions in the vicinity
of the GC including Sgr B1 and B2 inside the Sgr B Complex. 

Broadly, these authors' procedure is to 
determine the local energy density in non-thermal electrons required in order that 
these particles produce, via their Fe K-shell ionizing collisions, the observed
flux of Fe K$\alpha$ (6.4 keV) X-ray photons. The authors then set the local magnetic field strength to be
in equipartition with the energy density in this population and determine the synchrotron emission by the
electrons in this field. The magnetic fields arrived at in this way for Sgr B1 and B2 are
at the $\sim$ few $\times 10^{-5}$ G level. \citet{Yusef-Zadeh2006} then determine the inverse Compton (IC)
spectrum of photons up-scattered from mm wavelengths to TeV+ energies by the inferred high energy component
of the same electron population. This mechanism requires electrons with 
energies in excess of 30 TeV as remarked by \citet{Yusef-Zadeh2006}
themselves -- a severe challenge given that 
cooling (dominated by synchrotron emission) is so efficient at these energies
with
a loss timescale at 30 TeV of
only a few decades (for the expected magnetic field strengths given below).

We have repeated the analysis of \citet{Yusef-Zadeh2006} to confirm their findings.
Our detailed re-examination reveals, however,
that their models for Sgr~B1 and Sgr~B2 encounter a number of difficulties. 
Not the least of these 
is that, in the case of Sgr B2 at least, the model requires
an average magnetic field strength that is
0.045 mG, on the order of one order of magnitude less than actually
measured -- and set as a {\it lower limit} -- for this object by \citet{Crutcher1996}
(and also much less than the B field inferred on the basis of the Crutcher scaling, Eq.(\ref{eqn_Crutcher})).

Another problem for the Sgr~B2 model of \citet{Yusef-Zadeh2006} is that, in the average 
molecular hydrogen density assumed by them, $10^4 $ cm$^{-3}$,
their assumed power-law, primary electron distribution seems to produce too much
bremsstrahlung emission at both $\sim$ 100 MeV and $\sim$ TeV energies, surpassing the EGRET upper limit
and also the HESS data points (Yusef-Zadeh et al. actually invoked IC to explain the $\sim$ TeV data,
but bremsstrahlung will make an unavoidable -- and actually dominant --
contribution in this same energy range given that $\sim$30 TeV+
electrons are necessary in order that the IC mechanism work).
This objection might be dealt with, however, 
by invoking a somewhat reduced $n_{H_2} \simeq 5 \times 10^3$ cm$^{-3}$. 

Yet another problem, though, is in what seems to be a lack of self-consistency in the way these authors
deal with the cooling of the injected, primary electrons. In their calculation of the energy density in
their fitted, equipartition electron population, Yusef-Zadeh et al. assume a pure power-law
in electron {\it kinetic} energy between 10 keV and 1 GeV. Leaving aside the potential issue that
the assumption of a power law in kinetic energy (rather than momentum) is a rather unnatural one, the problem here is that
ionization loses at low energy should significantly distort an injected spectrum away from such power-law
behavior
(unless the time from injection of the particles is significantly less than the ionization cooling timescale, $\ll$ 100 yr). 
Indeed, cooling of the distribution assumed by \citet{Yusef-Zadeh2006} would seem to be necessary in order
that bremsstrahlung emission by lower emergy electrons not over-produce continuum X-rays -- but then the general procedure 
of these authors would seem to lack self-consistency as, on the one hand, the X-ray emission is apparently
calculated with the cooled
distribution but the total energy density in the distribution is calculated with an uncooled (i.e., pure power law) distribution.

\subsection{A New Leptonic Model for Sgr B}

Because of these difficulties, and to further explore the parameter space,
we therefore relax the constraint imposed by \citet{Yusef-Zadeh2006} 
that the energy density in the electron population be the same as in the
magnetic field.
We arrive at a satisfactory, alternative
lepton scenario to that proposed by these authors
in the following fashion: adopting the expected values $n_{H_2} = 10^4$ cm$^{-3}$,
$B = 10^{-4}$ Gauss and $N_{H_2} = 8 \times 10^{23}$ cm$^{-2}$, we find the {\it cooled} electron distribution (as parameterized
by injection \spin and normalization) that reproduces the required $\zeta_{CR}$ and roughly reproduces the continuum
X-ray spectrum of Sgr B (via bremsstrahlung). 
We find the spectral index at injection mustbe close to $\gamma = 2.85$.

Fixing $\gamma$ and also the overall electron population normalization to the values determined by the procedure above,
we then determine the B field required in order that the electron spectrum found above reproduce the 330 MHz VLA datum. 
A very 
reasonable field of 1.3 $10^{-4}$ G is necessary
here for the case of the minimum possible Sgr B mass (the maximum mass case is excluded because it produces significantly
too much
$\sim$ GeV bremsstrahlung emission contravening the EGRET limit).
We subsequently check that the new value of B does not substantially alter $\zeta_{CR}$ and the X-ray flux for the already-determined
\spin  and normalization.

We then check the $\sim$ GeV bremsstrahlung emission from this distribution (I.C. is sub-dominant at this energy)
in the expected $\langle n_{H_2} \rangle$. 
Finding that the spectrum now exceeds the EGRET limit, we dial the $\langle n_{H_2} \rangle$ downwards until
Sgr B just escapes detection. The required $\langle n_{H_2} \rangle$ is $5 \times 10^3$ cm$^{-3}$. 
This is very close to the critical number density established by Eq.(\ref{eqn_critDnsty})
for molecular gas at a distance of $r_{\rm Sgr \, B} \sim 100 pc$ from the actual GC, viz., 
$\sim 6 \times 10^3$ cm$^{-3}$.

Again, we determine whether the $\langle n_{H_2} \rangle$ substantially alters $\zeta_{CR}$ and the X-ray flux for
the previously determined \spin  and normalization. We find that $\zeta_{CR}$ is largely unaltered, but the X-ray flux has
now decreased. We then vary the final degree of freedom open to us to adjust, the average column density through the cloud
and the ISM to the emission region. A satisfactory fit can be determined by dialling this down to be $4 \times 10^{23}$ cm$^{-2}$ 
-- a value well within the phenomenologically-allowable range. 

We have now fully constrained the system and can {\it predict}  
the 843 MHz flux. This prediction is 22.3 Jansky, 
fully consistent with the poorly-determined
observational value of 20 $\pm 10$ Jy.
Given that estimated fluxes might, {\it a priori},
vary by orders of magnitude, this represents good agreement.

Note that through this procedure we have arrived at an allowable parameter set that is probably not unique
and, given the complexities and non-negligible uncertainties surrounding the input data, we have not attempted a
$\chi^2$ analysis.
Non-trivially, however, we have found a one-zone model for the Sgr B complex that reproduces the broad features
of its low-energy, non-thermal phenomenology: a single electron population, injected as a power law in momentum 
with reasonable spectral index and
then cooled by the processes of ionization, bremsstrahlung and synchrotron radiation into a steady state distribution 
will reproduce the observed cosmic ray ionization rate, the 330 MHz datum and
the broad features of the continuum X-ray emission, all for very reasonable values of the ambient magnetic field,
molecular hydrogen number density and column density.

We note that the energy density in the cooled electron distribution is 2 eV cm$^{-3}$. 
This is around 10 times the energy density in CR electrons through the Galactic disk
as determined by \citet{Webber1998}, but
considerably sub-equipartition
with respect to 
the energy density in the fitted 1.3 $10^{-4}$ G magnetic field, viz. $\sim$ 400 eV cm$^{-3}$.
Note, however, that
the energy density represented by turbulent motions of the Sgr B gas
could easily be
in equipartition with such a field. 
In fact, adopting 
a line width of $10 $km s$^{-1}$, as detected for the envelope of Sgr B2 \citep{Lis1989}, e.g., 
one determines an equipartition field of $\sim$ 0.7 mG \citep{Novak1997}.
One also notes that the inferred energy density is considerably less than that found
for the Sgr B1 and B2 regions in the analysis of \citet{Yusef-Zadeh2006}, 21 and 51 eV cm$^{-3}$, respectively.
On the other hand it is considerably in excess of the energy density in the 
broad scale, non-thermal electron population recently inferred by
\citet{LaRosa2005}. These authors have, on the basis of the application of an
argument 
of equipartition (between magnetic field energy density and relativistic particles)
to observations of a large-scale ($6^\circ \times 2^\circ$),
diffuse flux of non-thermal radio emission (detected at 74 and 330 MHz) across the GC region, inferred 
an average magnetic field strength through this region of $\sim$ 15 $\mu$G and
a non-thermal electron energy density of $\sim 0.06$ eV cm$^{-3}$ (assuming
an electron energy density 1/100 that in protons).


The cooled electron distribution also represents a total energy of $1.1 \times 10^{48}$ erg. 
The energy required, however, to have been injected with the un-cooled initial spectrum --
and since lost mainly into ionization (thus heating) of the cloud matter -- is $1.4 \times 10^{50}$ erg.

The one respect in which our leptonic model can be said to fail is at TeV energies: it produces far too little
bremsstrahlung and IC emission in this energy range to be able to explain the HESS observations. On the other
hand it is certainly not in conflict with these observations.

\section{A Mixed Model for the Broadband Emission of Sgr B}

\begin{figure}
\plotone{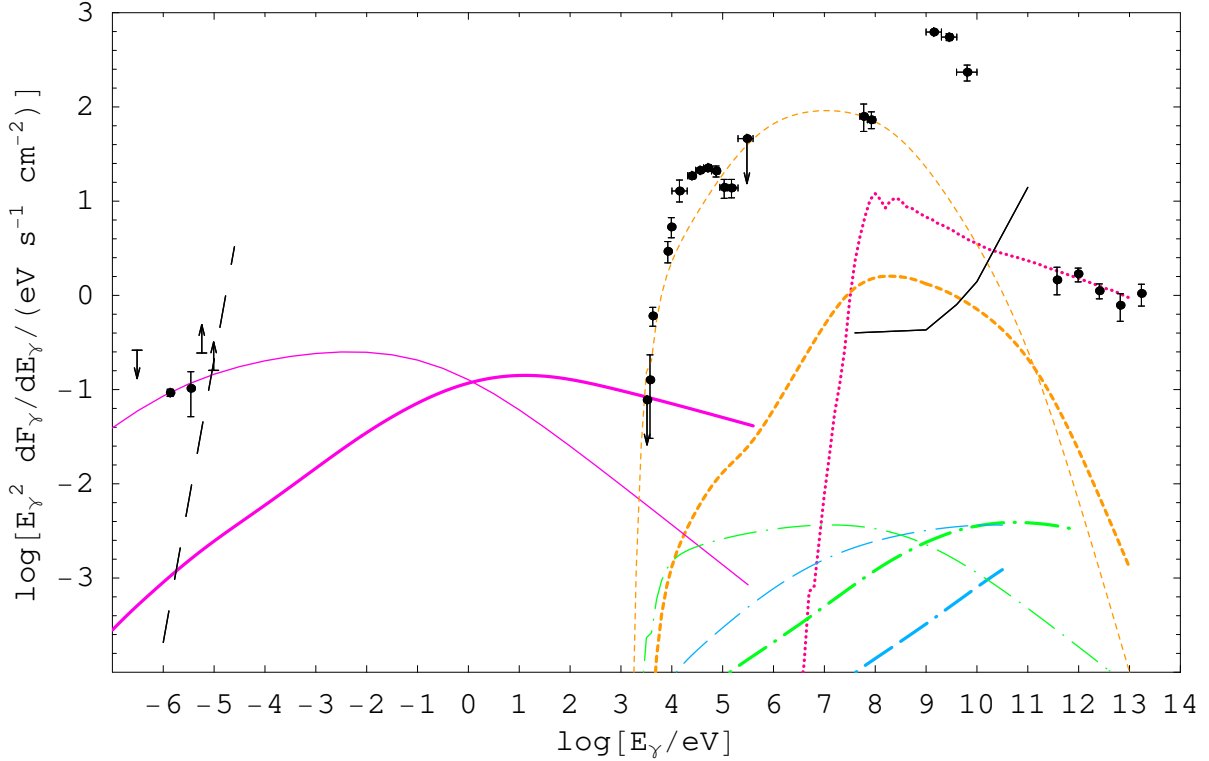}
\caption{A hybrid model for the broadband emission of the Sgr B Complex invoking
neutral meson production in CRH collisions and subsequent decay to explain the observed
TeV gamma rays and a primary electron distribution to explain the remainder of the
pertinent phenomenology. Processes
corresponding to various colored curves are as described in fig. \ref{fig_plotH2aBB}, thick
curves correspond to emission by secondary particles created in hadronic collisions
and thin curves to emission by primary leptons
(the dashed black curve is thermal bremsstrahlung emission as described in fig.(\ref{fig_plotH2aBB})).
}
\label{fig_plotMxdBB}
\end{figure}

We are finally, then, led to consider a hybrid model in which a relatively steep population (at injection) 
of low energy electrons 
is responsible for the bulk of the measured ionization rate, the 330 and 843 MHz flux and the X-ray 
continuum flux (via bremsstrahlung), 
whereas a hard population of CRH's dominates the emission at
TeV energies producing the bulk of gamma rays at this energy through neutral pion production and decay.
For the broad-band spectrum of such a mixed model, please see Figure \ref{fig_plotMxdBB}.

\section{Discussion}

Both of our models -- the double power law CRH model and the mixed lepton/hadron model --
require a hard, high-energy (TeV+) spectrum of hadrons to explain the HESS $\gamma$-rays.
An IC/primary lepton model for this emission -- at least as suggested by \citet{Yusef-Zadeh2006} -- does
not seem to work.
As initially noted in \citet{Aharonian2006} and as subsequently explored in 
\citet{Busching2006} and \citet{Ballantyne2007}, that the spectral index
of the diffuse $\gamma$-ray emission detected by HESS 
does not seem to vary over the extent of the CMZ, and, moreover,
is so similar to that detected for the central point-like source coincident with Sgr A*,
supports the notion that the CRH population responsible for this emission
has its origins at relatively small distances from the central black hole and diffuses out
into the CMZ from this position.
The total energy in this population has been estimated 
to be $10^{50}$ erg \citep{Aharonian2006} marginally consistent
with an origin in a single supernova.
Furthermore, adopting a diffusion coefficient
appropriate to cosmic ray diffusion through the Galactic disk, viz.
$D \sim 10^{30}$ cm$^{2}$ s$^{-1}$ at several TeV,
\citet{Aharonian2006} note that the break-down of the correlation between 
$\gamma$-ray emission and molecular density at angular scales of $\sim$ 1.3$^\circ$
from the actual GC (a position to the East of Sgr B) implies a definite time
for the injection event of around 10 000 years, perhaps consistent with the unusual SNR
Sgr A East being the source of the CRH population \citep{Crocker2005}.

There are a number of caveats here, however.
Firstly, 
one might expect in such a scenario that 
-- at the periphery of the diffuse emission region (just before the breakdown of
the $\gamma$-ray/molecular density correlation) -- the spectral index of the emission should harden, but such an effect is
not seen.
Furthermore, the adoption of a diffusion coefficient appropriate to the Galactic disk carries with it
a large uncertainty (as noted by \citet{Aharonian2006}).
Certainly, if there is a coherent global GC magnetic field of $\sim$ mG strength, then, at the least,
the timing of the assumed injection event has to be blow out substantially, and in fact,
the whole diffusion-away-from-central-source picture may not be tenable \citep{Morris2007}.

A steady-state picture for the origin of these CRHs -- though one still assuming a central source -- may be
attractive, therefore. One such model that has recently been investigated in the context 
of stochastic acceleration on the turbulent magnetic field close to the central black hole \citep{Liu2006}
is that of \citet{Ballantyne2007}.

\subsection{Origin of Putative Steep, Low-Energy Hadron Population}

As described at length above, one model that reproduces the broadband phenomenology
of the Sgr B Complex consists of a double power law CRH population, the steeper power-law
becoming dominant at low energy. 
Despite the modest CRH energy density required in this scenario ($\sim 2$ eV cm$^{-3}$), 
one should certainly question the naturalness of such an overall spectrum: if the 
CRHs are assumed to originate outside the cloud and to be diffusing into it, then exactly
the opposite sort of spectral behavior as outlined would be expected, viz., a progressive
flattening towards low energies as lower-energy particles find it increasingly difficult 
to diffuse into the cloud
before catastrophic energy loss via hadronic collision on ambient matter in the cloud:
see \citet{Gabici2006} and also the companion paper to the current work \citep{Jones2007}.

On the other hand, if the CRHs are accelerated {\it inside} the GMC, 
it may be 
that 
lower energy CRHs are unable to leave the cloud and naturally accumulate, thereby steepening the
overall spectrum towards lower energies, precisely as we require in our model.
So, in this scenario, a hard, CRH population 
-- perhaps originating outside the cloud but
with sufficient rigidity to penetrate the cloud at high energies
-- is dominant at HESS energies whereas a population accelerated in situ, 
and perhaps trapped by the cloud's turbulent magnetic fields,
is dominant at lower energies and responsible for most of the Sgr B phenomenology we report.

The total energy in the double power law CRH spectrum is between 1.3-3.3 $\times 10^{48}$ erg, a small
fraction of the mechanical energy available from a SN explosion, e.g..

Another question hangs over the magnetic field required in
the double power law hadronic scenario. This is, at $\sim$2-4 mG,
possessed of an energy density 400-1600 times larger than that of
the ``expected" field given by the Crutcher scaling as noted above.
This scaling relation, however, does not seem to  describe the situation in
Sgr B2 particularly well, as also already noted. 
Furthermore, as stressed, the direct Zeeman splitting determination of the Sgr B2 field, at $\sim$ 0.5 mG, 
actually defines a lower limit to the local average field sgtrength.
In fact, both \citet{Lis1989} and
\citet{Crutcher1996} actually countenance average magnetic field strengths as high as $\sim$2 mG
for the Sgr B2 cloud on the basis of the theoretical prejudice that the cloud be magnetically
supported against gravitational collapse -- 
so we cannot exclude that such field strengths may actually apply on large scales in the complex.

Finally, it must also be admitted that the
large magnetic field required 
may be an artefact of our assumption of a 1-zone model: 
if p's are accelerated inside the cloud then they will naturally accumulate more
in regions where B fields are higher and the synchrotron emissivity resulting
from secondaries created by collisions in these regions will also be relatively higher
in these locations
-- so the assumption of an ``average" value
for the B field may not adequately reflect these effects. 

\subsection{Origin of Putative Electron Population}

In terms of fitting to the broad-band spectrum, the other successful model discovered above
is one that invokes a steep ($E^{-2.9}$) primary lepton spectrum at injection together with a hard
CRH population to explain the HESS $\gamma$-rays.

One notes that 
the required (steep) electron injection index required in this scenario
is well beyond the range normally attributed to shock acceleration,
which is typically 2.0-2.4 and, indeed, even a combination of
shocks might not allow such an overall steep distribution easily.
There do seem, however, to be two plausible instances in which
such a steep spectrum might be arrived at:
(i) stochastic acceleration off a turbulent
magnetic field within the cloud (i.e., acceleration by plasma wave turbulence,
a second-order Fermi acceleration process; see, e.g. \cite{Petrosian2004}), and (ii)
shock acceleration with an energy-dependent loss or diffusion.
Again, concrete examples of these two general mechanisms will be
evaluated in a later work \citep{Crocker2006}.

The energy at injection
represented by the putative non-thermal electron population invoked above, $1.4 \times 10^{50}$ erg,  
is too big, e.g., to have been supplied by a single, ordinary supernova event (the total energy 
that goes into non-thermal particles populations and fields being around $5 \times 10^{49}$ erg; see,
e.g. \cite{Duric1995} -- but, of non-thermal particles, most energy goes into hadrons).

In this context, however, one notes the possibility that multiple supernovae have occurred
within the Sgr B Complex, it being a site of very active star formation and harboring many massive and hot stars.
In fact, \citet{Koyama2006a} and \citet{Koyama2006b} have very recently claimed the discovery of a new SNR 
designated Suzaku J1747.0-2824.5 (G0.61+0.01) on the basis of a detection
of an excess of 6.7 keV FeXXV K$\alpha$ emission from inside the Sgr B region  with
the X-ray Imaging Spectrometer on board the orbitting {\it Suzaku}
X-ray telescope.
Furthermore, on the basis of the CO mapping performed by \citet{Oka1998}, \citet{Koyama2006a} and \citet{Koyama2006b}
claim the existence of a radio shell in this same region
possessed of a kinetic energy of order $10^{52}$ erg, supporting the possibility of multiple
supernovae within the Sgr B Complex.

\section{Conclusion}

A major result of this work is that we get far
too little radio flux from secondary leptons (normalized to the
HESS $\gamma$-ray data)
to explain the VLA and SUMSS observations for reasonable magnetic field values
and assuming a power-law (in momentum) behavior of the initiating CRH primaries.
Furthermore --
making the (perhaps naive) assumption that the CRHs be governed by a single power law 
in momentum from ultra-relativistic energies down to the sub-relativistic regime --
the cosmic ray ionization rate $\zeta_{CR}$ implied by the CRH population
inferred from the HESS $\gamma$-ray data is far too small to be reconciled with 
recent determinations of this quantity for Sgr B2. 
Conversely, a simple interpretation of the ionization rate in
this cloud being maintined solely by the CRH distribution would
then overproduce TeV photons via pp scattering events compared
to what is measured by HESS (again, assuming a pure power law).

Another major result is that it seems almost certain that one needs a hard, high-energy cosmic-ray {\it hadron}
population to explain the HESS TeV+ $\gamma$-ray data: a particular model introduced by \citet{Yusef-Zadeh2006}
that would seek to explain this emission by inverse Compton scattering by a population of
primary cosmic ray electrons seems to break down when considered in detail. 
Certainly, as for the comsic ray hadron case,
a single power-law distribution of primary electrons cannot account for all Sgr B phenomenology.

We have investigated two scenarios that can be reconciled with all the data:
\begin{enumerate}
\item A scenario invoking two CRH
power law populations (with a steep spectrum dominant at low energy). This scenario requires an ambient
magnetic field in the Sgr B Complex in the range 2-4 mG that, while apparently not excluded by existing
Zeeman splitting and polarimetry data, may be uncomfortably high. 
Future submillimetre polarimetry measurements 
may soon rule in or rule out the necessary field strength.
Zeeman splitting measurements at mm wavelengths with the next generation of instruments
such as the Atacama Large Millimeter Array (ALMA: \cite{Wootten2006}) 
may also have sufficient sensitivity to detect these high field strengths, even in lower density regions.
\item Following a path blazed by
\citet{Yusef-Zadeh2006}, we have investigated the idea that
primary electrons play a significant role and, in fact, explain the bulk
of the non-thermal Sgr B phenomenology. Unfortunately we find that
the particular instantiation of a leptonic model arrived at by 
\citet{Yusef-Zadeh2006} does not seem to be phenomenologically viable (requiring as it does
too small a magnetic field to be reconcilable with the data) or self-consistent (in its treatment of
spectral distortion due to ionization cooling at low energies).
We have found, however, a leptonic scenario which does satisfactorily account for much of the phenomenology of Sgr B
(viz., the CR ionization rate, the X-ray continuum emission, and the 330 and 843 MHz radio emission)  
but a hadronic component, contrary to the opinions of \citet{Yusef-Zadeh2006}, 
is apparently necessary to explain the gamma-ray emission as mentioned above.

\end{enumerate}

\section{Acknowledgements}
The authors thank Jim Hinton for providing the HESS data in numerical form
and Crystal Brogan for supplying the 74 and 330 MHz data VLA data in numerical form.
RMC thanks Roger Clay
for discussions about cosmic ray diffusion, Gavin Rowell for advice about the analysis of the $\gamma$-ray 
data, and Troy Porter for advice about the psf of EGRET.

\clearpage

\section{Appendix A: Average Parent Proton Energies of Secondaries}

\begin{deluxetable}{lcccccc}
\tablewidth{0pt}
\tablecaption{The average parent CRH energy of a $\gamma$-ray observed at $E_\gamma$ expressed as 
a multiple of $E_\gamma$ for various spectral indices.\label{table_parentEnergyGammaRays}}
\tablehead{
  \colhead{}
& \colhead{2.0}
& \colhead{2.2}
& \colhead{2.4}
& \colhead{2.6}
& \colhead{2.8}
& \colhead{3.0}

}
\startdata

$10^7$ eV    &11212.1   &3307.03   &1492.45   &906.29    &642.835   &494.906  \\
$10^8$ eV    &381.042   &93.5398   &39.2354   &24.2151   &18.2917   &15.2974  \\
$10^9$ eV    &402.378   &124.055   &52.8439   &29.721    &20.0141   &15.0221  \\
$10^{10}$ eV &347.006   &136.529   &64.3183   &36.7628   &24.5029   &18.111  \\
$10^{11}$ eV &196.61    &95.9581   &49.9323   &28.7306   &18.4379   &13.0247  \\
$10^{12}$ eV &90.378    &59.6075   &39.3149   &26.5121   &18.5825   &13.6492  \\
$10^{13}$ eV &28.9777   &24.4022   &20.311    &16.792    &13.8674   &11.5044  

\enddata
\end{deluxetable}

\begin{deluxetable}{lcccccc}
\tablewidth{0pt}
\tablecaption{The average parent CRH energy of a secondary electron observed at $E_e$ expressed as 
a multiple of $E_e$ for various spectral indices.\label{table_parentEnergyElectrons}}
\tablehead{
  \colhead{}
& \colhead{2.0}
& \colhead{2.2}
& \colhead{2.4}
& \colhead{2.6}
& \colhead{2.8}
& \colhead{3.0}

}
\startdata

$10^7$ eV    &20509.2 &   4435.49 &   1315.68 &   548.547 &   305.384 &   209.977   \\
$10^8$ eV    &839.363 &   226.115 &   95.561 &    56.3558 &   40.0325 &   31.4774  \\
$10^9$ eV    &647.868 &   206.759 &   88.6688 &   49.7365 &   33.3554 &   24.8667  \\
$10^{10}$ eV &647.707 &   260.024 &   117.842 &   62.529 &    38.4941 &   26.6127   \\
$10^{11}$ eV &362.318 &   194.075 &   106.836 &   62.655 &    39.8574 &   27.4842 \\
$10^{12}$ eV &143.603 &   103.072 &   73.0004 &   51.8339 &   37.439 &    27.8002  \\
$10^{13}$ eV &39.3796 &   35.013 &    30.7712 &   26.7642 &   23.0849 &   19.7976   

\enddata
\end{deluxetable}

\begin{deluxetable}{lcccccc}
\tablewidth{0pt}
\tablecaption{The average parent CRH energy of a secondary positron observed at $E_e$ expressed as 
a multiple of $E_e$ for various spectral indices.\label{table_parentEnergyPositrons}}
\tablehead{
  \colhead{}
& \colhead{2.0}
& \colhead{2.2}
& \colhead{2.4}
& \colhead{2.6}
& \colhead{2.8}
& \colhead{3.0}

}
\startdata

$10^7$ eV    &4711.5 &    999.145 &   368.44 &    216.004 &   163.045 &   138.591 \\
$10^8$ eV    &441.384 &   114.894 &   51.1813 &   32.9168 &   25.4181 &   21.4796  \\
$10^9$ eV    &489.759 &   153.778 &   66.9793 &   38.7002 &   26.6889 &   20.347 \\
$10^{10}$ eV &525.323 &   205.124 &   92.4019 &   49.6369 &   31.1947 &   22.01    \\
$10^{11}$ eV &305.33 &    160.058 &   87.2869 &   51.3001 &   32.9662 &   23.0505 \\
$10^{12}$ eV &126.953 &   89.2382 &   62.0621 &   43.4609 &   31.1185 &   23.0179   \\
$10^{13}$ eV &36.8475 &   32.5944 &   28.563 &    24.8512 &   21.5276 &   18.6259    

\enddata
\end{deluxetable}

\clearpage


\clearpage

\end{document}